\documentclass[lettersize,journal]{IEEEtran}
\usepackage{amsmath,amsfonts,amssymb}
\usepackage{algorithmic}
\usepackage{algorithm}
\usepackage{array}
\usepackage[caption=false,font=normalsize,labelfont=sf,textfont=sf]{subfig}
\usepackage{textcomp}
\usepackage{stfloats}
\usepackage{url}
\usepackage{verbatim}
\usepackage{graphicx}
\usepackage{cite}
\usepackage{mathtools}

\usepackage{diagbox}
\usepackage{orcidlink}
\usepackage{bbm}
\usepackage{gensymb}

\usepackage{hhline}

\usepackage{nameref}
\newcounter{mylabelcounter}

\makeatletter
\newcommand{\labelText}[2]{%
#1\refstepcounter{mylabelcounter}%
\immediate\write\@auxout{%
  \string\newlabel{#2}{{1}{\thepage}{{\unexpanded{#1}}}{mylabelcounter.\number\value{mylabelcounter}}{}}%
}%
}
\makeatother

\begin{document}

\title{TinyML for On-Device and Edge Analytics in Wireless Networks: A Survey of Deployments, Opportunities, and Concept-Drift Mitigation}

\author{Prasoon Raghuwanshi\orcidlink{0000-0002-9629-9742},~\IEEEmembership{Student Member,~IEEE},
Vimal Bhatia\orcidlink{0000-0001-5148-6643},~\IEEEmembership{Senior Member,~IEEE},
Sridhar Iyer\orcidlink{0000-0002-8466-3316},~\IEEEmembership{Senior Member,~IEEE},
Matti Latva-aho\orcidlink{0000-0002-6261-0969},~\IEEEmembership{Fellow,~IEEE},
Onel Luis Alcaraz López\orcidlink{0000-0003-1838-5183},~\IEEEmembership{Senior Member,~IEEE},
%\vspace{-10mm}
\thanks{Prasoon Raghuwanshi, Onel López, and Matti Latva-aho are with the Centre for Wireless Communications, University of Oulu, $90570$, Oulu, Finland (e-mail: Prasoon.Raghuwanshi@oulu.fi; Onel.AlcarazLopez@oulu.fi; Matti.Latva-aho@oulu.fi).}
\thanks{Vimal Bhatia is with the Department of Electrical Engineering, Indian Institute of Technology Indore, $453552$, Indore, India, with the Skoda Auto University, $29301$, Mlada Boleslav, Czech Republic, with the Faculty of Informatics and Management, University of Hradec Krolove, $50003$, Hradec Krolove, Czechia, and with the University of Oulu, $90570$, Oulu, Finland (e-mail: vbhatia@iiti.ac.in)}
\thanks{Sridhar Iyer is with the Department of ECE, S.G. Balekundri Institute of Technology, Belagavi, Karnataka, $590010$, India (e-mail: sridhariyer@sgbit.edu.in)}
\thanks{This research has been supported by the Research Council of Finland (Grants 362782 (ECO-LITE), and 369116 (6G Flagship)), the European Commission through the Horizon Europe/JU SNS project Ambient-6G (Grant 101192113), the Riitta ja Jorma J. Takasen säätiö (Grant 20240358), the Nokia Scholarship (Grant 20260695), and the Oulun yliopiston tukisäätiö (Grant 20260126).}}



\maketitle

\begin{abstract}
Ubiquitous intelligence is essential for enabling real-time, adaptive, autonomous, and scalable operations in the next generation of wireless networks.
However, this poses significant challenges in data management and energy consumption on the end-device/edge side, specially under dynamic environmental conditions.
This has driven the adoption of tiny machine learning (tinyML), which offers data-driven optimization at the end-device/edge side.
In this work, we survey and thoroughly discuss various tapped/untapped deployment possibilities of tinyML in wireless networks.
We identify existing frameworks, accustomed to design tinyML algorithms, that could be utilized to solve a range of wireless network problems.
We present a federated learning-based tinyML model update procedure, for both battery-powered and batteryless end-devices, to resolve the concept drift problem faced by tinyML models.
Furthermore, we discuss the update-aware checkpointing, fault-tolerant bootloader, and intermittent-aware modify operation, which could support federated learning-based tinyML model update in the case of batteryless end-devices.
Overall, this paper spells out several areas where end-device/edge intelligence can be utilized in the next generation of wireless systems, as well as ways to mitigate the concept drift problem faced in the case of end-device intelligence.

\end{abstract}


\section{Introduction}

\begin{figure*}[!t]
\centering
\includegraphics[width=\linewidth]{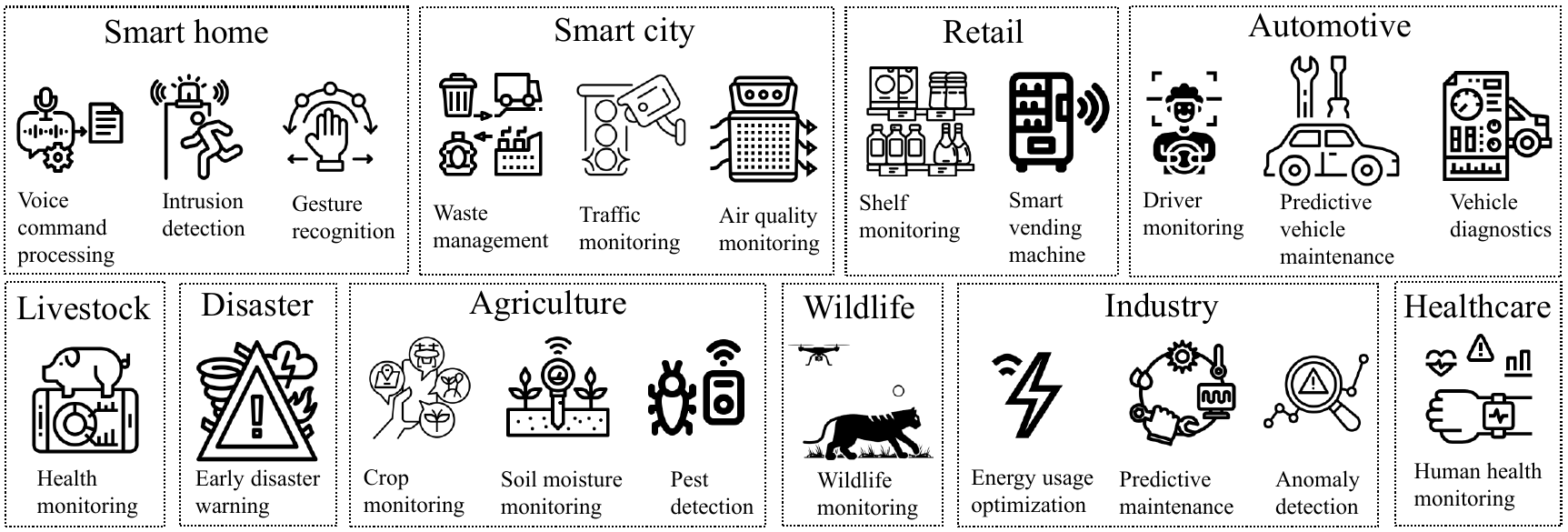}
\caption{Several key tinyML use cases and specific applications.}
\label{generalApplicationsFigure}
\end{figure*}

\IEEEPARstart{T}{iny} machine learning (tinyML) is a transformative technology that enables machine learning (ML) to be deployed on microcontroller (MCU)-based edge nodes and end-devices like Internet of Things (IoT) devices.
This allows end-devices to perform inference directly on the onboard sensor's data, and edge nodes on aggregated data.
Currently, there are approximately $250$ billion MCUs in use worldwide, and this number is expected to grow by an astonishing $40$ billion annually over the next decade \cite{10.1145/3608473}.
As a result, tinyML is poised to become a ubiquitous technology in the near future, with several applications illustrated in Fig.~\ref{generalApplicationsFigure}.
In $2022$ alone, the number of tinyML device installations sits around $2$ billion, and this number is projected to reach $11$ billion in $2027$.\footnote{\url{https://go.abiresearch.com/lp-37-technology-stats-you-need-to-know-for-2023}}
TinyML not only supports sustainable computing but also has a significantly smaller carbon footprint compared to traditional cloud-based systems.
This positions tinyML as a key enabling technology in achieving the United Nations' energy sustainability goals \cite{10285066}.

TinyML can be applied to various tasks within wireless communication networks. For instance, it can be used to intelligently perform wireless channel estimation and data symbol detection at the receiver (RX) side \cite{9745826}, pre-distortion at the transmitter (TX) side in a multiple-input multiple-output (MIMO) orthogonal frequency-division multiplexing (OFDM) system \cite{10242459}, radio frequency (RF)-based indoor localization \cite{avellaneda2023tinyml, 10060797}, on-device sensor linearization \cite{BOTEROVALENCIA2023e00477}, etc.
Notably, tinyML is expected to contribute significantly to achieve \textit{energy efficiency} and \textit{sustainability}, which are respectively key performance and value indicators for the sixth-generation (6G) of wireless communication networks \cite{10317778, shiraishi2023energy}.

In the case of IoT systems, onboard analytics of sensor output/data, powered mainly by tinyML, can be performed using a fraction of the computational resources required by the cloud-based data analysis \cite{10.1145/3608473}.
Onboard inference not only enhances data privacy and reduces unnecessary wireless data transmissions, but also eliminates the need for continuous connectivity between end-devices and central entities, thereby minimizing reliance on centralized infrastructure \cite{10284551, bonneau2023addressing}.
Moreover, tinyML significantly reduces the energy cost associated with ML algorithms by lowering their memory requirements to just a few hundred kilobytes, utilizing reduced-precision arithmetic,  removing redundant arithmetic operations, and decreasing inference time \cite{10284551}.
Regarding IoT sensing, tinyML can help perform this function with lower-quality sensors, such as low-resolution cameras, hence reducing the overall IoT cost/footprint \cite{10.1145/3608473}.
However, whether to perform sensed data analysis locally onboard or on the cloud, by offloading the data to it, is not an easy decision, as it depends on the computational cost of the tinyML-based data analysis task and the cost of data offloading.
The former can be quantified in terms of required energy, memory, and processing power, while for the latter, it is the required transceiver energy, bandwidth, and latency \cite{10285066}.

\subsection{Motivation and Key Contributions}

A summary of the state-of-the-art tinyML-related survey papers, from $2022$ onwards, is available in Table~\ref{surveyPapers_table}.
After reviewing this state-of-the-art literature, we realized that the recently published survey papers are focusing on either one or a combination of the following themes:
(i) ML model compression \cite{10177729, 1011453661820, 1011453657282, 10433185, le2023efficient, RAY20221595, 9912325, info14080470},
(ii) hardware platforms \cite{fi14120363, 10177729, 10433185, electronics13173562, RAY20221595, info14080470}, inference engines \cite{fi14120363, 10177729, kallimani2024tinyml, 1011453661820, 10433185, electronics13173562, le2023efficient, RAY20221595, 9912325}, hardware accelerators \cite{1011453657282, 10433185, RAY20221595}, dataflow and data locality optimization \cite{1011453657282, info14080470},
(iii) existing categories of tinyML algorithms \cite{10433185},
(iv) challenges faced by tinyML technology \cite{fi14120363, 10177729, kallimani2024tinyml, 1011453661820, electronics13173562, le2023efficient, RAY20221595, 9912325, 11557456}, and
(v) applications in computer vision \cite{fi14120363, kallimani2024tinyml, 1011453657282, electronics13173562, a17110476, 10970940, RAY20221595, 9912325}, healthcare \cite{fi14120363, 10177729, kallimani2024tinyml, 1011453661820, s25103191, KEIVANIMEHR2025109653, electronics13173562, RAY20221595, 9912325}, human behavior analysis \cite{10979983}, predictive maintenance \cite{10781407}, mobile robotics \cite{a17110476}, automotive \cite{fi14120363, 1011453661820, s25103191, electronics13173562, RAY20221595, 11557456}, agriculture \cite{fi14120363, 10177729, 1011453661820, s25103191, electronics13173562, RAY20221595}, speech recognition \cite{fi14120363, kallimani2024tinyml, electronics13173562, RAY20221595, 9912325}, and anomaly detection \cite{fi14120363, 10177729, kallimani2024tinyml, RAY20221595, 9912325}.
However, none of the existing survey works comprehensively discuss the deployment possibilities of tinyML in wireless networks to address a broad range of wireless communication problems.
In particular, the existing survey works overlook the integration of tinyML with battery-powered end-devices, batteryless end-devices, and low-power edge nodes.
No attention has been paid to identifying potentially unexplored domains for tinyML in wireless networks.
The existing survey works also overlook the emerging role of federated learning (FL)-assisted tinyML model updates in wireless environments, especially for intermittently powered batteryless devices where issues such as unreliable energy availability, checkpointing, and fault-tolerant model updates become critical.
Therefore, a dedicated survey focusing on tinyML-driven wireless networks is both timely and essential to bridge these gaps in this rapidly emerging interdisciplinary domain.

This work completely focuses on the utilization of tinyML in wireless networks, and its specific contributions are as follows:
\begin{itemize}
    \item We overview the existing tinyML algorithms suitable for battery-powered/batteryless end-devices and low-power edge nodes, as well as their respective use cases in wireless networks.
    \item We survey various purposes for which tinyML-based end-device/edge inference has already been employed in the existing wireless network literature.
    \item We survey and discuss several new research domains in wireless networks where the integration of tinyML-based end-device/edge inference could offer significant benefits.
    \item We present and discuss the FL-based tinyML model update procedure for both battery-powered and batteryless end-devices.
    Moreover, we discuss the intermittent-aware modify operation, update-aware checkpointing, and fault-tolerant bootloader, which can support FL-based tinyML model update in the case of batteryless end-device.
\end{itemize}

{
\setlength\arrayrulewidth{1pt}
\begin{table*}[t]
\caption{State-of-the-Art Survey Papers on TinyML\label{surveyPapers_table}}
\centering
\begin{tabular}{@{}p{0.55cm} l p{16.2cm}@{}}
\hline
\textbf{Ref.} & \textbf{Year} & \textbf{Key focus} \\ 
\hline
\cite{RAY20221595} & $2022$ & $\bullet$ Overview of ML model compression techniques, hardware platforms, hardware accelerators, and inference engines preferred for tinyML \newline
$\bullet$ Review of the application of tinyML in speech recognition, object detection/classification, healthcare, automotive, agriculture, and anomaly detection, as well as challenges faced by tinyML technology \\
\hline
\cite{9912325} & $2022$ & $\bullet$ Overview of ML model compression techniques, inference engines preferred for tinyML, data engineering techniques to collect input data for tinyML models, and feature projection techniques to extract features from the aforementioned input data \newline
$\bullet$ Review of the application of tinyML in image recognition, anomaly detection, speech recognition, and healthcare, as well as challenges faced by tinyML technology \\
\hline
\cite{fi14120363} & $2022$ & $\bullet$ Overview of hardware platforms and inference engines preferred for tinyML deployment, as well as challenges faced by tinyML technology \newline
$\bullet$ Review of benefits of tinyML in object detection/classification, anomaly detection, agriculture, automotive, healthcare, speech recognition \\
\hline
\cite{10177729} & $2023$ & $\bullet$ Overview of ML model compression techniques, as well as hardware platforms and inference engines preferred for tinyML \newline
$\bullet$ Review of benefits of tinyML in healthcare, smart farming, and anomaly detection, as well as challenges faced by tinyML technology \\
\hline
\cite{info14080470} & $2023$ & $\bullet$ Overview of model compression techniques, hardware platforms, and dataflow optimization for tiny deep neural network models \\
\hline
\cite{1011453661820} & $2024$ & $\bullet$ Overview of ML model compression techniques and inference engines preferred for the deployment of tinyML model on MCU \newline
$\bullet$ Review of benefits of tinyML in healthcare, automotive, agriculture, and industry as well as challenges faced by tinyML technology  \\
\hline
\cite{1011453657282} & $2024$ & $\bullet$ Discussion on compression techniques for deep learning (DL) models as well as the existing tiny DL models for vision-based applications \newline
$\bullet$ Discussion on hardware accelerators, dataflow types, and data locality optimization mechanisms for tiny DL models \\
\hline
\cite{10433185} & $2024$ & $\bullet$ Overview of ML model compression techniques and existing categories of tinyML algorithms such as supervised, unsupervised, reinforcement, self-supervised, weakly supervised, meta, and continual   \newline
$\bullet$ Discussion on hardware accelerators, hardware platforms, and inference engines preferred for tinyML  \\
\hline
\cite{electronics13173562} & $2024$ & $\bullet$ Overview of hardware platforms and inference engines preferred for tinyML deployment, as well as challenges faced by tinyML technology  \newline
$\bullet$ Review of the application of tinyML in speech recognition, object detection/classification, healthcare, automotive, agriculture, environmental monitoring, and surveillance  \\
\hline
\cite{kallimani2024tinyml} & $2024$ & $\bullet$ Overview of existing tinyML frameworks and inference engines, as well as challenges faced by tinyML technology \newline
$\bullet$ Review of benefits of tinyML in speech recognition, object detection/classification, healthcare, phenomics, and anomaly detection  \\
\hline
\cite{a17110476} & $2024$ & $\bullet$ Discussion on tinyML in mobile robotics and image processing tasks \\
\hline
\cite{10781407} & $2024$ & $\bullet$ Discussion on tinyML in predictive maintenance  \\
\hline
\cite{10970940} & $2025$ & $\bullet$ Discussion on the utilization of tinyML for unmanned aerial vehicle-based computer vision tasks such as vegetation segmentation, object detection, forest fire detection, etc  \\
\hline
\cite{s25103191} & $2025$ & $\bullet$ Discussion on the application of tinyML in healthcare, automotive, and agriculture  \\
\hline
\cite{KEIVANIMEHR2025109653} & $2025$ & $\bullet$ Discussion on the potential of tinyML in identifying cardiovascular diseases by facilitating pervasive cardiovascular monitoring \\
\hline
\cite{10979983} & $2025$ & $\bullet$ Discussion on tinyML in human behavior analysis  \\
\hline
\cite{le2023efficient} & $2026$ & $\bullet$ Overview of ML model compression techniques, inference engines preferred for tinyML deployment, and challenges faced by tinyML in image classification/recognition and speech recognition tasks \\
\hline
\cite{11557456} & $2026$ & $\bullet$ Discussion on the application of tinyML in transportation systems, and challenges faced in it \\
\hline
This work & $2026$ & $\bullet$ Discussion on the existing tinyML algorithms suitable for the battery-powered/batteryless end-device and low-power edge node as well as their respective use case in wireless networks \newline
$\bullet$ Review of the existing literature on tinyML-based end-device/edge inference in wireless network \newline
$\bullet$ Discussion on the untapped deployment possibilities of tinyML-based end-device/edge inference in wireless networks  \newline
$\bullet$ Discussion on the tinyML model update procedures, for both battery-powered and batteryless end-devices, to deal with concept drift  \\
\hline
\end{tabular}
\end{table*}
}

\subsection{Research Methodology and Organization}

To ensure methodological rigor, well-defined inclusion and exclusion criteria were established for identifying relevant studies.
Empirical studies that explicitly implemented tinyML or lightweight ML on MCU-based end-devices and edge nodes within the context of wireless networks and those, either theoretical or implementation-based, revealing potentially unexplored domains for tinyML in wireless networks were included, while those relying solely on cloud-based implementation were excluded.

The literature search was conducted using a comprehensive set of keywords, including wireless networks, IoT, tinyML, MCU, on-device inference, embedded ML, battery-powered end-device, batteryless end-device, and low-power edge node.
To ensure broad coverage, the search was performed across multiple scientific databases, including IEEE Xplore, Elsevier, Springer, MDPI, and Google Scholar.
A systematic review of the literature identified $146$ candidate studies, distributed across four domains: $18$ state-of-the-art survey papers on tinyML, $42$ works on tinyML frameworks for low-power end-devices and edge nodes, $28$ and $35$ works on already tapped and untapped deployment possibilities of tinyML in wireless networks, and $12$ works dealing with concept drift in tinyML.
For the selection of these works, titles and abstracts were inspected first to filter out those out-of-scope, while the remaining papers underwent full-text eligibility assessment against the aforementioned inclusion and exclusion criteria.

The rest of the paper is organized as follows.
Section~\ref{ML_Algos} discusses the existing tinyML algorithms suitable for the end-device/edge node.
Section~\ref{tinyMLexistingDeplymentPossibilities} overviews the utilization of tinyML in the existing wireless network literature.
Section~\ref{UntappedDeplymentPossibilities} identifies untapped deployment possibilities of tinyML in wireless networks.
Section~\ref{dealConceptDrift} discusses the FL-based tinyML model update procedure.
Finally, we conclude our work in Section~\ref{conclUSion} and provide key challenges and research directions with respect to the protocol aspects for integration of tinyML to 6G wireless networks.

{
\setlength\arrayrulewidth{1pt}
\begin{table*}[!t]
\caption{Overview of ML Model Compression Methods\label{ML_CompressionMethod_table}}
\centering
\begin{tabular}{@{}p{1.5cm} p{8cm} p{3cm} p{4.5cm}@{}}
\hline
\textbf{Method} & \textbf{Explanation} & \textbf{Advantages} & \textbf{Disadvantages} \\ 
\hline
Knowledge distillation & $\bullet$ Knowledge distillation is an approach to train a compact ML model, known as the student model, with the help of a large and accurate ML model, known as the teacher model, such that the student model's output predictions can mimic the teacher model's output predictions as closely as possible  \newline
$\bullet$ The compactness of the student model is defined in terms of the number of trainable parameters \newline
$\bullet$ For training the student model,
(i) output predictions of the teacher model are used as the target, and
(ii) the distillation loss function is used, which takes into account both the error in the student model's output predictions and their similarity to the aforementioned target
& $\bullet$ Accelerates model inference as the student model performs fewer operations than the teacher model
& $\bullet$ Require careful tuning of the distillation loss function hyperparameters, as these hyperparameters control the teacher-to-student knowledge transfer \newline
$\bullet$ Large training time as model training has to be performed at least twice \\
\hline
Pruning & $\bullet$ Pruning is a method to trim an ML model by eliminating its non-critical weight parameters \newline
$\bullet$ There are mainly two methods for pruning: structured and unstructured \newline
$\bullet$ The structured method eliminates weights in the form of blocks, such as an entire row/column of a weight matrix, based on a specific redundancy criterion \newline
$\bullet$ The unstructured method eliminates an individual weight based on its absolute value
& $\bullet$ The structured method accelerates model inference on hardware \newline
$\bullet$ The unstructured method can compress the model to any degree
& $\bullet$ The structured method reduces model accuracy \newline
$\bullet$ The unstructured method creates irregular model structures, which lead to operations involving sparse matrices during model inference.
Meanwhile, the presence of sparse matrices de-accelerates model inference on most hardware
\\
\hline
Quantization & $\bullet$ Quantization is a method that maps the ML model parameters lying within a large set, often a continuous range with 64/32-bit precision, to a smaller set, often discrete with 8-bit or lower precision \newline
$\bullet$ Quantization can be performed during or after the ML model's training phase.
The former is called quantization-aware training, while the latter is post-training quantization \newline
$\bullet$ Post-training quantization is widely adopted in the case of 8-bit precision settings \newline
$\bullet$ Quantization-aware training is a superior option compared to post-training quantization in the case of below 8-bit precision settings
& $\bullet$ Model accuracy loss in the case of quantization is relatively less than pruning
& $\bullet$ The quantization function \cite{le2023efficient} is non-differentiable and can result in zero gradients for low-bit precisions.
This makes it difficult to train a model with quantization-aware training  \\
\hline
Low-rank matrix decomposition & $\bullet$ Low-rank matrix decomposition is a method to approximate a high-rank weight matrix with a product of two low-rank matrices.
This minimizes the weight matrix's dimensionality, ultimately leading to a compact ML model \newline
& $\bullet$ Reduces model size by removing redundant weight parameters.
Thus, highly effective in the case of models with a large number of redundant parameters
& $\bullet$ Additional analyses are needed to find the optimal rank for the product matrices and for hyperparameter tuning.
Moreover, these analyses do not generalize for every task/application \newline
$\bullet$ No significant acceleration in the model inference  \\
\hline
\end{tabular}
\end{table*}
}

\section{TinyMLs for Low-Power End and Edge Devices}\label{ML_Algos}

To transform a standard neural network (NN)-based ML model into a tinyML model for MCU-based low-power end-devices and edge nodes, several techniques are employed, including knowledge distillation, pruning, quantization, and low-rank matrix decomposition \cite{le2023efficient}.
These methods, overviewed in Table~\ref{ML_CompressionMethod_table}, carry out this transformation by scaling the ML model according to the available resources on the MCU.
In parallel, numerous frameworks have been developed by technology companies to facilitate the implementation of tinyML models on MCU-based devices.
Notable examples include TensorFlow Lite,\footnote{\url{https://www.tensorflow.org/lite}} embedded learning library,\footnote{\url{https://microsoft.github.io/ELL/}} ARM-NN,\footnote{\url{https://github.com/ARM-software/armnn}} CMSIS-NN,\footnote{\url{https://arm-software.github.io/CMSIS_5/NN/html/}} and STM32Cube.AI,\footnote{\url{https://stm32ai.st.com/stm32-cube-ai/}} and others.

TinyML is often associated with compressed NN-based ML, but many classical ML methods are also suited for MCU-scale deployment because they naturally satisfy strict constraints on memory, latency, and energy \cite{lopez2024zero}.
Classical ML methods such as decision trees \cite{11175363, 11488232}, random forest \cite{11175363, 11488232, 11361023}, and gradient-boosted trees \cite{11175363} can be used, since inference in them consists primarily of simple branching logic rather than matrix multiplications, making them computationally inexpensive.
Similarly, linear regression \cite{11488232}, logistic regression \cite{11488232}, and linear support vector machines \cite{11361023} are attractive because they reduce inference to dot products that can be efficiently implemented with fixed-point arithmetic.
Probabilistic methods, like Naive Bayes \cite{11488232}, are particularly lightweight, relying only on stored probabilities and simple arithmetic.
Besides, there are also methods to directly form tinyML models without first training a large ML model and then compressing it.
One major approach is constraint-aware or hardware-aware training, such as the AutoTinyML framework \cite{10298625} for decision trees and random forest, where memory usage, inference latency, and energy consumption are incorporated into the optimization objective during training, ensuring the resulting model is deployable from the outset.
Another pathway involves using libraries such as emlearn,\footnote{\url{https://github.com/emlearn/emlearn}} which directly generate optimized C code for classical ML models, or platforms like Edge Impulse,\footnote{\url{https://www.edgeimpulse.com/}} which allow developers to design full pipelines, including feature extraction and lightweight classifiers, specifically for embedded targets.
Finally, specialized algorithms designed explicitly for edge environments, such as Bonsai \cite{9531793} or ProtoNN \cite{9531793}, demonstrate that tinyML models can be constructed natively for constrained hardware rather than derived from larger models.

Next, we discuss existing tinyML algorithms/frameworks designed for end-devices and low-power edge nodes, along with their use cases in wireless networks.
Table~\ref{tinyML_algorithms_table} summarizes these algorithms/frameworks.

{
\setlength\arrayrulewidth{1pt}
\begin{table*}[t]
\caption{TinyML Algorithms for Battery-Powered/Batteryless End-Devices and Low-Power Edge Node \label{tinyML_algorithms_table}}
\centering
\begin{minipage}{\linewidth}
\begin{tabular}{@{}p{2.9cm} p{2.6cm} p{0.4cm} p{1.6cm}| p{0.01cm} @{}p{2.9cm} p{2.6cm} p{0.4cm} p{1.6cm}@{}}
\hline 
\textbf{TinyML Framework} & \textbf{Target Board/Processor} & \textbf{Alg. Type} & \textbf{Task} & & \textbf{TinyML Framework} & \textbf{Target Board/Processor} & \textbf{Alg. Type} & \textbf{Task} \\
\hline
\end{tabular}
\end{minipage}\vfill
\vspace{-0.1mm}
\begin{minipage}{0.49\linewidth}
\begin{tabular}{@{}p{2.9cm} p{2.6cm} p{0.4cm} p{1.6cm}|}
Memory-aware hybrid \cite{9516678} & MSP430F552x boards & DL & Classification  \\
TyCNN \cite{9892925} & ESP32 board & DL & Classification  \\
Differentiable neural architecture search \cite{MLSYS2021_c4d41d96} & STM32 boards & DL & Classification  \\
Randomized matrix approximation \cite{9115654, JMLR_v14_wang13c} & STM32 and ESP32 boards & DL & General  \\
Reservoir sampling-cum-local outliner factor \cite{szydlo2022online} & Arduino Nano 33 BLE board & UL & Classification  \\
Extreme value theory \cite{10261398} & Raspberry Pi Pico & UL & Classification  \\
AutoCloud k-fix \cite{10219125} & STM32 and ESP32 boards & UL & Classification  \\
TinyCleanEDF \cite{fi16020042} & Raspberry Pi 4 & UL & Feature extraction  \\
CompressEdgeML \cite{fi16020042} & Raspberry Pi 4 & DL & Data compression  \\
OnceNAS \cite{ZHANG2024120567} & Raspberry Pi 4 & DL & Classification  \\
Weight factorization \cite{10187678} & ESP32 board & DL & Classification  \\
Modified VGG16 \cite{9502691} & Raspberry Pi 3, Raspberry Pi 4 & DL & Classification  \\
Modified LeNet-5 \cite{9502691} & Raspberry Pi 3, Raspberry Pi 4 & DL & Classification  \\
PoPS \cite{8962235} & STM32 and ESP32 boards & RL & Sequential decision-making 
\end{tabular}
\end{minipage}\hfill
\begin{minipage}{0.49\linewidth}
\begin{tabular}{@{}p{2.9cm} p{2.6cm} p{0.4cm} p{1.6cm}@{}}
TinyNS \cite{10_1145_3603171} & STM32 boards & DL & Classification  \\
Resource scalers-cum-optimization tuner \cite{svoboda2020resource} & ESP32 board, STM32 board with ARM Cortex-M0  & RL & Sequential decision-making \\
\vspace{-2.0mm} tinyMAN \cite{osti_10334227} & \vspace{-2.0mm} TI CC2652R board, STM32 board with ARM Cortex-M4 & \vspace{-2.0mm} RL & \vspace{-2.0mm} Sequential decision-making \\
Genesis \cite{10_1145_3297858_3304011} & & DL & Classification  \\
LBPNet \cite{9093550} & STM32 and ESP32 boards & DL & Classification  \\
Training-cum-implementation-cum-checkpointing \cite{9774756} & MSP430FR599x boards & DL & Classification  \\
Distilled pruning \cite{10570946} & STM32 boards & DL & AMC  \\
Distilled quantization \cite{10570946} & STM32 boards & DL & AMC  \\
MobileNetV2 \cite{9969601} & STM32 boards & DL & Classification  \\
TinyM$^2$Net-V3 \cite{rashid2024tinymnetv} & Raspberry Pi 4 & DL & Classification  \\
SqueezeNet \cite{9969601} & STM32 boards & DL & Classification  \\
SquishedNets \cite{shafiee2017squishednets} & STM32 boards & DL & Classification  \\
MCUNet \cite{NEURIPS2020_86c51678} & ARM Cortex-M4, Cortex-M7, STM32 boards  & DL & Classification  \\
EtinyNet \cite{Xu_Li_Zhang_Lai_Gu_2022} & STM32 boards & DL & Classification, Feature extraction 
\end{tabular}
\end{minipage}\vfill
\vspace{-3.5mm}
\begin{minipage}{\linewidth}
\begin{tabular}{p{0.97\linewidth}}
\\ \hline
\end{tabular}
\end{minipage}
\end{table*}
}

\begin{figure*}[!t]
\centering
\includegraphics[width=\linewidth]{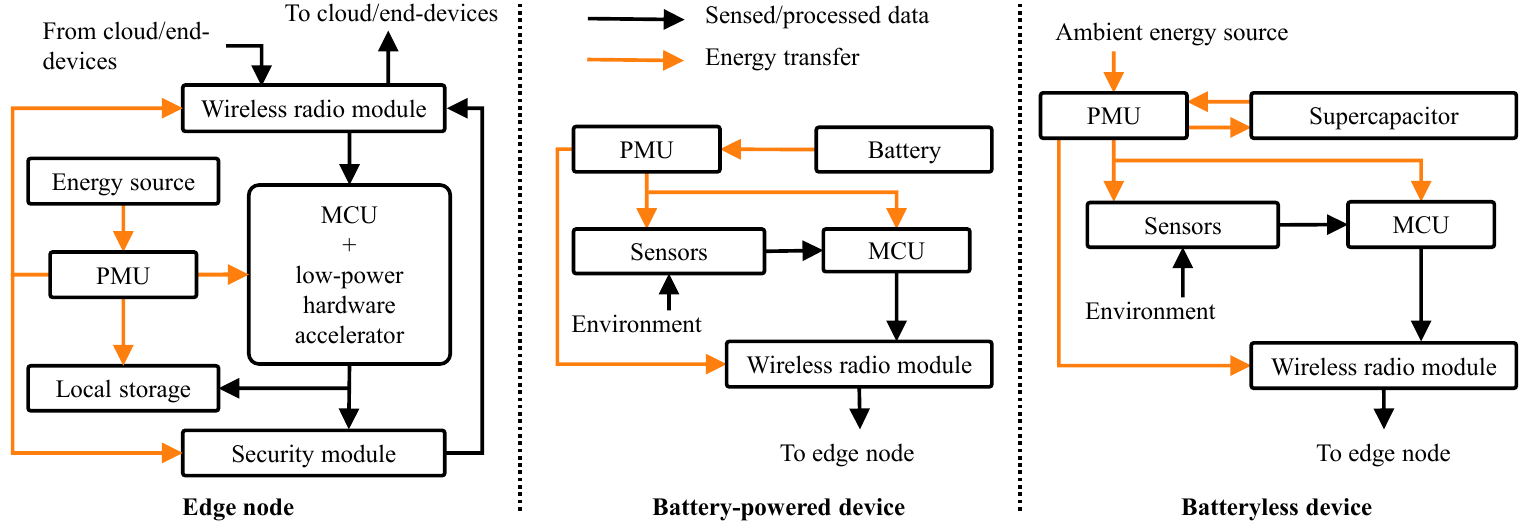}
\caption{Hardware architecture of a low-power edge node, battery-powered end-device, and batteryless end-device.}
\label{deviceArchitecturesFigure}
\end{figure*}

\subsection{Low-Power Edge Node}\label{ML_edgeAlgos}

Low-power edge nodes represent a broader class of embedded systems designed to perform data processing close to the data source, while operating under resource constraints.
These nodes are commonly deployed for tasks where real-time decision-making and reduced communication latency are critical \cite{KULANDAIVEL2026101963}.
The hardware architecture of low-power edge nodes typically consists of an energy source, a power management unit (PMU), a wireless radio module, a low-power MCU, a low-power hardware accelerator, local storage, and a security module, as illustrated in Fig.~\ref{deviceArchitecturesFigure}.
The energy source may consist of mains power or a battery, depending on deployment requirements.
The energy source feeds the PMU that regulates the voltage supplied to edge node components, monitors the energy source's state-of-charge, and manages the duty cycle, i.e., transitions between active and sleep modes, of the edge node.
The wireless radio module is often integrated with multiple communication interfaces, such as southbound and northbound interfaces \cite{s26010314}.
This allows the edge node to aggregate data from heterogeneous end-device networks, provide interoperability among end-device networks using different communication standards, forward aggregated data to the cloud, receive control commands from the cloud/controller and forward them either to an actuator or end-devices, etc \cite{8489908}.
The MCU, along with a low-power hardware accelerator, performs data filtering, aggregation, and local analytics.
Data filtering removes noise, invalid readings, duplicates, and irrelevant information before further processing.
Data aggregation combines multiple measurements into summarized forms such as averages, minimums, maximums, statistical distributions, or event counts.
Data analytics then applies algorithms ranging from simple threshold-based detection to tinyML inference.
Local storage enables temporary buffering of sensor data, caching of ML models, maintenance of system logs, and operation during intermittent upstream network connectivity.
The security module implements authentication and encryption, ensuring the integrity and confidentiality of transmitted information.
Meanwhile, the architectural characteristics of low-power edge nodes influence tinyML model designs/choices by looking at how compact and optimized the models must be, hence balancing accuracy, latency, and energy usage.
Next, we provide existing tinyML algorithms/frameworks implementable on low-power edge nodes.

Tiny deep learning (DL) models built upon the OnceNAS \cite{ZHANG2024120567} framework as well as those compressed with respect to the weight factorization \cite{10187678} technique, can be used for \textit{cooperative spectrum sensing} (CSS) task, \textit{automatic modulation classification} (AMC) task, \textit{specific emitter identification} (SEI) task, \textit{unintended load detection} (ULD) task in wireless power transfer (WPT) systems, and TX-coil \textit{activation/deactivation} task in a multi-TX WPT system.
OnceNAS explores the DL model search space by mapping candidate models with their respective characteristics, including parameter count and inference latency/accuracy.
Based on this mapping, OnceNAS chooses the model favorable for the target edge node.
Weight factorization technique operates in an iterative manner, where it factorizes the weight matrix of a layer into two matrices to minimize the weight matrix reconstruction error as well as the number of parameters and floating point operations of the respective DL model's layer.
Tiny DL models built upon the OnceNAS framework and weight factorization technique are also credible for wireless \textit{channel estimation}, \textit{RF fingerprinting-based positioning}, \textit{data symbol detection}, \textit{intrusion detection} in network traffic, RX device's \textit{automatic gain control (AGC) index range} prediction where edge node is the RX, and the following estimation tasks on the TX side of the WPT system:
\textit{compensation capacitance}, RX's \textit{load resistance}, \textit{power} delivered to RX's load resistor, and \textit{coupling coefficient} between the TX and RX sides.

Compressed versions of VTCNN2 \cite{10570946}, ResNet \cite{10570946}, and InceptionNet \cite{10570946} models with respect to the distilled pruning \cite{10570946} and distilled quantization \cite{10570946} methods can be used for \textit{AMC} task.
Distilled pruning and distilled quantization follow the following procedure.
First, both implement knowledge distillation to obtain a lightweight distilled model.
Next, distilled pruning implements the Net-trim \cite{10570946}, an unstructured pruning technique, to maximize the sparsity of weights present in the distilled model's layers without affecting the distilled model's performance.
Meanwhile, distilled quantization implements the product quantization \cite{10570946} algorithm, which transforms the weight matrix of the distilled model's layers into their respective product quantization codes.
These codes reduce the precision of weights as well as compress the weight matrix, as the ratio of the memory required to store the original weight matrix to its respective product quantization codes is greater than $1$.

Reinforcement learning (RL) algorithms built upon the PoPS \cite{8962235} and resource scalers-cum-optimization tuner \cite{svoboda2020resource} framework are supported by a low-power edge node and can be used to perform the IoT \textit{device scheduling} task in environment monitoring/control scenarios and the task to maintain \textit{constant power} at RX's load resistor in a WPT system.
PoPS leverages transfer learning and pruning to train and reduce redundancy of the deep RL (DRL) model.
The resource scalers-cum-optimization tuner framework utilizes quantization, pruning, and neural architecture search techniques as the resource scalers, while the optimization tuner decides the combination of these resource scalers to be applied to a DRL model.

\subsection{Battery-Powered End-Devices}\label{ML_batteryPoweredAlgos}

Battery-powered end-devices are the most common class of embedded systems.
The hardware architecture of these devices typically consists of a battery, a PMU, sensors, an MCU, and a wireless radio module, as illustrated in Fig.~\ref{deviceArchitecturesFigure}.
The battery serves as the primary energy source.
The PMU continuously monitors the battery's state-of-charge, regulates voltage supplied to device components, and manages the duty cycle to maximize the operational lifetime of the device.
Sensors collect environmental data. These sensors are connected to a low-power MCU with clock speeds in the tens to hundreds of MHz range, random access memory (RAM) sizes from a few tens to a few hundreds kilobytes, and flash memory often under $1$ megabyte.
The MCU performs sensor data acquisition, filtering, local processing, and decision-making tasks.
Based on the MCU's decision, the wireless radio module transmits collected information to the edge node.
Since battery-powered end-devices operate on a finite energy supply and energy consumption directly affects battery lifetime, tinyML models for these devices must be computationally efficient and optimized to balance inference accuracy with energy overhead. Next, we provide existing tinyML algorithms/frameworks that meet the battery-powered device's constraints.

Tiny DL models built upon the memory-aware hybrid method \cite{9516678} can be used to perform \textit{proximity detection}, among mobile devices, stemmed upon received signal strength indicator (RSSI)-based ranging.
The memory-aware hybrid method optimizes the dataflow of DL models to minimize their energy consumption and latency caused by data movement.

The tiny DL models built upon the TinyNS \cite{10_1145_3603171} and TyCNN \cite{9892925} framework, as well as those designed by utilizing the differentiable neural architecture search \cite{MLSYS2021_c4d41d96} and randomized matrix approximation \cite{9115654, JMLR_v14_wang13c} compression technique, are credible for on-device \textit{anomaly detection} in the observed data.
TinyNS identifies the ideal combination of hyperparameters and operators for neurosymbolic AI \cite{10_1145_3603171} models, while TyCNN identifies the ideal number of ad-hoc dilated convolutional blocks \cite{9892925} for convolutional NN (CNN) models, ensuring compatibility with the resource constraints of the target device.
Differentiable neural architecture search samples an ML model, that satisfies the target device's resource constraints, from the given network search space.
Randomized matrix approximation technique replaces the sparse weight matrices of NN layers with small dense matrices, reducing both the memory required for weight matrix storage and the number of matrix-related computations.
Moreover, the tiny unsupervised learning (UL) algorithm based on the reservoir sampling-cum-local outliner factor framework \cite{szydlo2022online}, the extreme value theory \cite{10261398}, and the AutoCloud k-fix \cite{10219125} can also be used for the on-device \textit{anomaly detection} task.
Reservoir sampling-cum-local outliner factor framework leverages the concept of anomaly score \cite{szydlo2022online} to identify anomalous data.
Here, the local outliner factor algorithm computes the anomaly score of the new data point by taking into account the set of data points stored in the device's memory, while reservoir sampling selects the aforementioned set.
Extreme value theory offers a set of distributions to model extreme events, such as anomalies.
Once an appropriate distribution is chosen, it can be used to estimate the probability of an observed data point being an anomaly.
AutoCloud k-fix leverages the concept of typicity \cite{10219125} and eccentricity \cite{10219125} to identify anomalous data.
Observing vibrations of industrial rotating machines \cite{9533927} or electric motor bearing \cite{hu2025tin} are example scenarios for an anomaly detection task.

TinyML algorithm named TinyCleanEDF \cite{fi16020042} is credible for both on-device \textit{anomaly detection} and \textit{semantic encoding} tasks.
TinyCleanEDF employs FL models for anomaly detection and an autoencoder for semantic encoding.
CompressEdgeML \cite{fi16020042}, a tinyML algorithm, is well-suited for the \textit{semantic compression} task.
One of its key features is the ability to adapt its compression ratio following a change in available bandwidth and available memory in the device.
TinyCNN models built upon the EtinyNet \cite{Xu_Li_Zhang_Lai_Gu_2022} framework are credible for \textit{semantic encoding} task \cite{shiraishi2023energy}.
EtinyNet utilizes (i) the linear depthwise block \cite{Xu_Li_Zhang_Lai_Gu_2022} and its dense counterpart to reduce both the computations to be performed and parameters in the model, and (ii) adaptive scale quantization to minimize both the model size and accuracy drop due to quantization.

Meanwhile, Tiny DL models built upon the randomized matrix approximation compression technique can also be used for \textit{semantic encoding}, \textit{semantic compression}, on-device \textit{sensor linearization} \cite{Sensors_Excitation_Linearization}, \textit{data symbol detection}, on-device \textit{localization} stemmed upon two-way ranging features, on-device \textit{intrusion detection} in network traffic, and RX device's \textit{AGC index range} prediction tasks.
Besides, these tiny DL models can be used as the \textit{pre-distorter} on TX devices in a MIMO-OFDM system.

The tiny RL algorithm with continuous state/action space, such as proximal policy optimization (PPO) built upon the tinyMAN \cite{osti_10334227} framework, is useful for on-device energy management.
The memory footprint and energy consumption of tinyMAN-based RL algorithms are less than $100$KB and $\sim28\ \mu$J/inference, respectively \cite{osti_10334227}.
Moreover, the tiny RL algorithms built upon the PoPS and resource scalers-cum-optimization tuner framework can be used for on-device \textit{resource allocation}, for IoT devices operating in distributed settings, and transmission power management on Bluetooth low energy (BLE) beacons of a navigation system.

Tiny DL models namely MobileNetV2 \cite{9969601}, SqueezeNet \cite{9969601}, SquishedNets \cite{shafiee2017squishednets}, modified LeNet-5 \cite{9502691}, and modified VGG16 \cite{9502691}, as well as those built upon the TinyM$^2$Net-V3 \cite{rashid2024tinymnetv} and MCUNet \cite{NEURIPS2020_86c51678} framework can be used for histopathology \cite{10_1007_978_3_031_48121_5_17}, precision agriculture \cite{10_1007_978_3_031_48121_5_17}, product quality inspection/surveillance in industries \cite{10_1007_978_3_031_48121_5_17}, etc.
TinyM$^2$Net-V3 devises multimodel DL models and employs knowledge distillation as well as uniform 8-bit quantization to reduce their memory and computational requirements.
MobileNetV2 utilizes the depthwise separable convolution whose computational cost is $~8-9\times$ smaller than the standard convolution, while SqueezeNet utilizes both $1\times1$ and $3\times3$ convolutional filters with $1\times1$ being in majority, which effectively reduces the number of model parameters.
Along with this, implementing the model compression technique from \cite{9969601}, which involves decreasing the number of layers, pruning, and quantization, on MobileNetV2 and SqueezeNet further minimizes their memory footprint and makes them suitable for battery-powered devices.
SquishedNets is the compressed version of SqueezeNet and can be utilized for classification tasks comprising of less than $10$ classes.
Modified LeNet-5/VGG16, compressed version of LeNet-5/VGG16 models, utilizes node merging \cite{9502691} and automatic horizontal fusion \cite{9502691} to reduce inference latency.
MCUNet recognizes that expanding the model search space to include model architectures with higher floating-point operations per second, while adhering to the device’s resource constraints, can yield a better-performing model.
Building on this insight, MCUNet adopts the code generator-based compilation method \cite{NEURIPS2020_86c51678} to minimize memory overhead during inference.
This enables MCUNet to expand the model search space, allowing for the exploration of larger models.
Next, MCUNet optimizes the model search space as per the device's resource constraints and performs the neural architecture search within the optimized search space.

\subsection{Batteryless End-Devices}\label{ML_batterylessAlgos}

Batteryless end-devices eliminate conventional batteries and instead rely on energy harvesting for their energy feed.
The hardware architecture of these devices typically consists of an energy storage element (such as a supercapacitor), a PMU, sensors, an MCU, and a wireless radio module, as illustrated in Fig.~\ref{deviceArchitecturesFigure}.
In the case of batteryless devices, the PMU typically comprises an \textit{energy harvester} that harvests erratic electrical currents from ambient sources, an \textit{energy-harvesting interface} that conditions, rectifies, and regulates the harvested erratic electrical currents, an \textit{energy storage management logic} that controls charging and discharging of the energy storage element, and an \textit{energy-aware duty cycle control logic}.
Because of energy harvesting, batteryless devices experience intermittent energy availability.
As a result, the execution strategy of tinyML-based tasks on these devices must adapt to energy availability, choosing between local inference or offloading. TinyML models for these devices are pushed to the absolute minimal computational complexity. In some cases, multiple tinyML models of differing complexity might even be employed, dynamically selected based on energy budget, trading off accuracy for energy cost to maintain progress \cite{SABOVIC2023100736}. Next, we mention existing tinyML algorithms/frameworks implementable on batteryless end-devices.

The tiny DL model built upon the framework from \cite{10_1145_3417308_3430273} can be used for the face classification task with at most $5$ classes.
The tinyML model, named LBPNet \cite{9093550}, is viable for the object classification task.
LBPNet performs bit-shifting and bitwise-OR, a comparison operation, instead of computationally intense convolutions.
This allows LBPNet to have smaller memory/energy requirements and faster inference speed, relative to a CNN-based tiny model.
Next, tiny DL models built upon the framework from \cite{9774756} are viable for \textit{classification} tasks. 
The framework from \cite{9774756} utilizes (i) the block-circulant matrix-based model, structured pruning, and circular buffer \cite{9774756} to reduce model size, inference time, and inference memory/energy requirements, and (ii) the low-energy accelerator to perform vector operations \cite{9774756} required for implementation of the block-circulant matrix-based model.
Additionally, tiny DL models obtained after hyperparameter optimization performed by the Genesis \cite{10_1145_3297858_3304011}, a minimization tool for DL models, are supported by devices operating in the mW power range.

RL algorithms supporting the discrete state/action space could be implemented on batteryless end-devices.
When the cardinality of the discrete state/action space is low, Q-learning \cite{10012046} or state–action–reward–state–action (SARSA) \cite{10445292} algorithm may be preferred.
On the other hand, when the cardinality is high, Q-learning/SARSA with linear function approximation could be a viable option \cite{10445292}.
However, an RL algorithm supporting continuous state/action space, such as PPO \cite{schulman2017proximal}, actor-critic RL \cite{6392457}, meta RL \cite{10.1145/3486611.3486670}, etc, requires a large number of trial-and-error steps during its training phase to converge to an optimal RL policy.
This means training such RL algorithm would be both time- and energy-demanding.
Thus, it is not favorable to implement RL algorithms with continuous state/action space on batteryless end-devices, due to their inherent energy limitation.
Rather, an ideal option is to shift the learning to an edge node or remote server, as done in \cite{10012046, 10.1145/3486611.3486670, fi16120460}, and inform only the action to be taken to the batteryless end-device.

An ML model for multi-class classification demands training on large datasets, such that it can incorporate knowledge from large-scale data and gains the generality required for multi-class classification \cite{8611738}.
However, multi-class classification often leads to computationally expensive ML architectures.
Meanwhile, relative to the best average performing multi-class ML model, class-specific ML models demand a small amount of training data, have computationally inexpensive ML architectures, and achieve higher class-specific accuracy \cite{8611738}.
The reason is that class-specific ML models specialize in classifying a particular class rather than generalizing over multiple classes.
As a batteryless end-device has minuscule energy and computational capabilities, it makes sense to use it for single-class classification rather than multi-class classification, and class-specific ML models from \cite{8611738} can make this possible.
Furthermore, a cluster of batteryless end-devices, where each device is equipped with a different class-specific ML model to cover all the possible classes, can make multi-class classification possible even with batteryless end-devices.

\begin{figure*}[!t]
\centering
\includegraphics[width=0.75\linewidth]{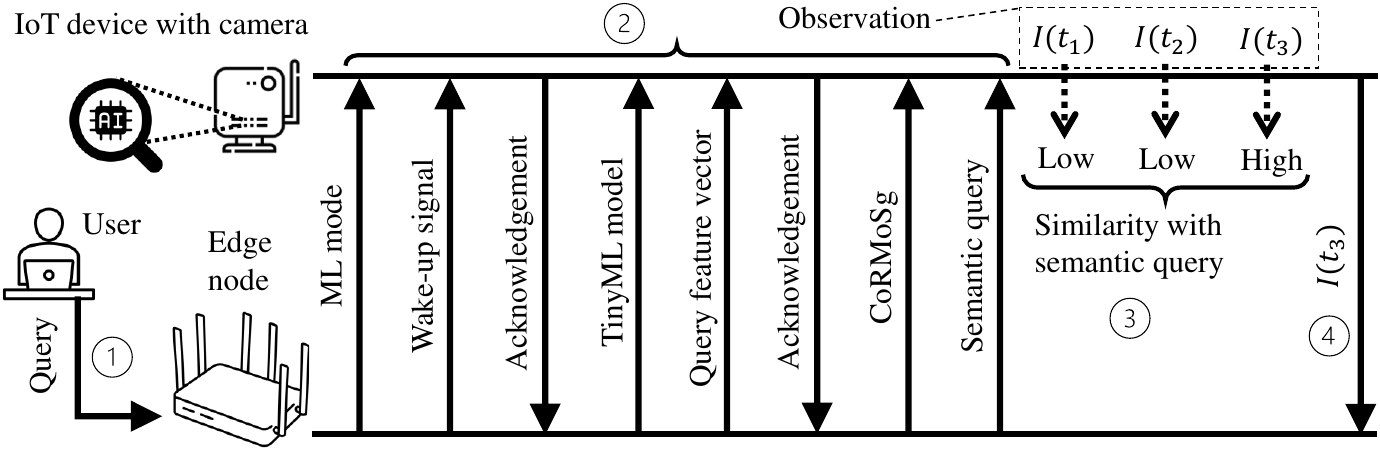}
\caption{TinyML-based image retrieval in a query-based IoT system.
The edge node receives a user query demanding an image with specific features.
Subsequently, the edge node broadcasts a mode-switching signal, termed ML-Mode, followed by the transmission of individual wake-up signals, tiny DL models, and query feature vectors.
Next, the edge node broadcasts a different mode-switching signal named CoRMoSg, followed by a semantic query.
Upon receiving CoRMoSg, the IoT devices process their observed image data using their respective tiny DL models, followed by similarity measure calculation based on the semantic query.
Thereupon, devices wake up their radios to transmit their respective images if the similarity measure of an image transcends the semantic query.}
\label{semanticDataTransmitFigure}
\end{figure*}

\subsection{Key Takeaways}

A key takeaway is that the architectural characteristics of end-devices and low-power edge nodes influence tinyML model designs/choices, impacting how compact and optimized the models must be and balancing accuracy and energy usage.
Both NN-based tinyML models and classical ML methods can be used for MCU-scale deployment.
Table~\ref{ML_CompressionMethod_table} overviews the techniques employed to scale standard NN-based ML models according to the available resources on MCUs.
Unlike edge nodes and battery-powered end-devices, it is not favorable to implement RL algorithms with continuous state/action spaces and multi-class classification algorithms on batteryless end-devices, due to the inherent energy limitation in them.
Instead, one may rely on single-class classification while shifting RL algorithms with continuous state/action space to an edge node or remote server that then may inform only the action to be taken to batteryless end-devices.
Lastly, existing tinyML frameworks can be utilized to perform different types of tasks in wireless networks, including classification, feature extraction, data compression, sequential decision-making, and estimation.
The summary of these tinyML frameworks is available in Table~\ref{tinyML_algorithms_table}.

\section{TinyML Deployment: Already Documented}\label{tinyMLexistingDeplymentPossibilities}

TinyML-based end-device/edge inference has several deployment possibilities in wireless communication networks, as outlined in Section~\ref{ML_Algos}.
Below, we survey those already documented in the literature.

\subsection{Semantically Relevant Data Transmission}

TinyML can be leveraged by an IoT device to make intelligent data transmission decisions based on the semantics of the data.
An example of this is the query-based image retrieval-cum-transmission task, as done in \cite{shiraishi2023energy}.
The procedure, illustrated in Fig.~\ref{semanticDataTransmitFigure}, flows as follows.
When the edge node receives a user query demanding a specific type of image, it initiates the ML model transmission phase.
This phase begins with the edge node broadcasting a mode-switching signal, termed ML-Mode, followed by the transmission of individual wake-up signals.
IoT devices respond to wake-up signals by transmitting the wake-up acknowledgment signals.
Subsequently, tiny DL models and query feature vectors are transmitted by the edge node to the IoT devices.
Here, transmitted tiny DL models are built upon the EtinyNet \cite{Xu_Li_Zhang_Lai_Gu_2022} framework.
Meanwhile, each model's size, in terms of the number of weights and biases, and the number of multiply-and-accumulation operations performed are $0.976$M and $117$M, respectively.
After the IoT devices acknowledge the signal to the edge node, they go back to sleep, which ends the ML model transmission phase.
Next, the edge node begins the image retrieval phase by broadcasting a different mode-switching signal, followed by the semantic query, which is the threshold similarity measure.
Upon receiving the mode-switching signal, the IoT devices start processing the observed image data using their respective tiny DL models.
The latter performs feature extraction followed by similarity measure calculation based on the semantic query.
The image retrieval phase ends with devices waking up their radios to transmit their respective images if the similarity measure of an image transcends the semantic query.
Note that the tiny DL model, used in \cite{shiraishi2023energy}, obtained an image retrieval accuracy in the range of $(0.5, 1]$ when the semantic query's magnitude decreases from $1$ to $0.6$.
Thus, this semantic-based data transmission approach prevents semantically non-relevant data transmissions.

\subsection{Smart On-Device Sensor Linearization}

A sensor's response to variations in a measured physical phenomenon can take various forms, such as a voltage/current signal.
Typically, this response exhibits a nonlinear relationship with the physical parameter being measured \cite{IslamMukhopadhyay_2019_1_21}.
For example, a small resistance change near 25$\degree$C in a thermistor represents a temperature change of only a few degrees, while the same resistance change near 0$\degree$C translates to a much greater temperature shift.
This means the same change in resistance does not correspond to the same change in temperature across the entire operating range.

With sensor linearization \cite{Sensors_Excitation_Linearization}, the response of a sensor can be scaled to measure the value of a physical parameter.
By applying sensor linearization, sensor responses become consistent, allowing the wireless sensor network (WSN) to produce high-quality datasets that better represent the actual variation in the measured physical parameter.
Sensor linearization simplifies the sensor calibration process, which is advantageous in the case of large WSNs \cite{IslamMukhopadhyay_2019_1_21}.
Sensor linearization also enhances network efficiency by reducing the need for repeated measurements, which minimizes unnecessary data transmissions.
Since data transmission dominates energy usage in WSNs, improving measurement quality at the sensor level contributes indirectly to extending network lifetime.
Since the sensor response typically has a nonlinear relationship with the physical parameter to be measured, polynomial functions often represent the linearization functions.
However, deriving an exact polynomial equation for the linearization function is a challenging task \cite{IslamMukhopadhyay_2019_1_21}.
Additionally, individual sensors, even of the same model, can exhibit unique manufacturing inconsistencies, temperature/environmental variations, and aging effects over time.
All of these factors uniquely influence the nonlinear relationship between the sensor's response and the physical parameter per sensor.

Tiny DL presents a viable solution to perform smart on-device sensor linearization on an MCU-based sensor-equipped device, as exemplified in \cite{BOTEROVALENCIA2023e00477}.
Therein, a tiny DL-based on-device sensor linearization was used to simultaneously linearize the responses of air, noise, and light pollution sensors used in a pollution measurement station.
Errors in the order of $\sim3.32$ Lux and $\sim4.0 \ \mu$g/m$^3$ in the case of light and air pollution measurements were respectively obtained.
In general, the tiny DL-based sensor linearization experienced an overall error of $\sim2.67\%$ in the case of measurements from all sensors.

\subsection{Process Radar Data}

Radar is an inevitable part of the perception sensor kit utilized in autonomous driving systems, where it is used for object detection and object heading estimation \cite{10186591}.
This is because radar-based perception is resilient to adverse climatic conditions, such as rain, haze, and snow, as well as to complex urban traffic scenarios \cite{s24092813}.
In current automotive radar systems, the object detection process begins with a fast Fourier transform applied to the raw radar data, converting it from the time domain data to the frequency domain data, known as radar images or fast Fourier transform maps \cite{10.1007/978-3-031-19839-7_23}.
These radar images are then processed using the constant false alarm rate algorithm \cite{iet:/content/books/ra/sbra021e} to generate radar points, also referred to as point cloud data \cite{s24092813}.
These radar points are subsequently used for object detection.
However, it is important to note that the significant amount of the target characteristic information available in raw radar data is lost in the case of radar points, whereas this information stays intact in the radar images.

Recently, there has been a growing interest among researchers in leveraging CNN to harness the rich contextual information available in radar images.
This approach has the potential to enhance object detection, classification, and heading estimation of both big and small objects \cite{10186591, s24092813, 10.1007/978-3-031-19839-7_23}.
However, as noted in \cite{9249018}, the number of parameters in a CNN-based radar object recognition model can scale up to $106$M, leading to memory demands that exceed the capacity of MCU-based automotive radar systems, motivating the use of tinyML variants.
An example of such a tinyML-cum-radar-based system is available in \cite{9892925}, where tiny CNN models are built upon the TyCNN framework for the indoor and in-car presence detection tasks.
As for indoor presence detection, the tiny CNN model's size, the number of multiply-and-accumulation operations it performs, and the accuracy it achieved are $33.2$kB, $4.2$M, and $99.5\%$, respectively.
Meanwhile, in the case of in-car presence detection, values for aforementioned parameters are $56.6$kB, $9.2$M, and $90.6\%$, respectively.

\subsection{RF Fingerprinting-Based Positioning}\label{RF_fingerprinting_positioning}

\begin{figure}[!t]
\centering
\includegraphics[width=\linewidth]{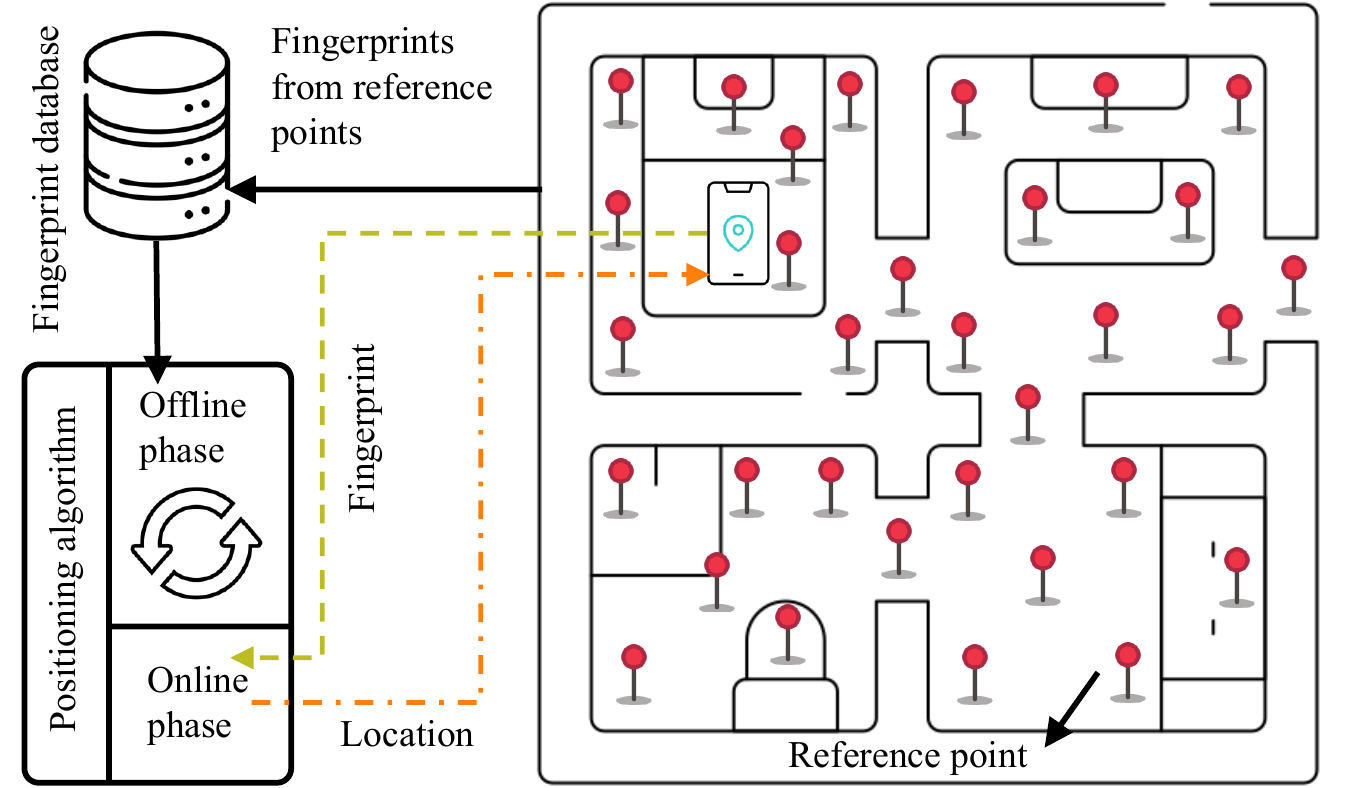}
\caption{Illustration of RF fingerprinting-based positioning. First, in the offline phase, fingerprints from various reference points are collected and stored in the fingerprint database, which is then used to train the positioning algorithm. Next, in the online phase, the trained positioning algorithm infers the asset's location given its fingerprint.}
\label{RFfingerprintingfigure}
\end{figure}

RF fingerprinting-based positioning method estimates the spatial location of an asset using a unique, location-dependent feature known as a fingerprint.
Typically, the \textit{received signal strength} of the signal transmitted by the TX attached to the asset acts as a fingerprint.
An edge node or existing radio technology infrastructure, such as a WiFi access point, BLE beacon, and cellular base station, acts as the receiving end of the aforementioned transmitted signal.
Depending on the computational requirements, RF fingerprinting-based positioning could either be performed on the edge node or the cloud server.
The aforementioned radio technology infrastructure relays fingerprints to the cloud server.

RF fingerprinting positioning operates in two phases, as illustrated in Fig.~\ref{RFfingerprintingfigure}.
First, \textit{offline phase}, where fingerprints from various reference points within the area of interest are collected and stored in a database.
This database is then used to train a positioning algorithm, which learns to map a fingerprint to the reference point's spatial coordinates or a sub-region where those spatial coordinates are likely to be found.
Second, \textit{online phase}, where the trained positioning algorithm infers the asset's location given its fingerprint.
RF fingerprinting-based positioning is particularly suitable for indoor positioning, as it can operate effectively in non-line-of-sight settings \cite{alhomayani2020deep}.

Recently, tiny DL has become a popular choice for positioning algorithms because of its following features \cite{alhomayani2020deep}:
(i) autonomous feature extraction from the fingerprint,
(ii) ability to form non-linear boundaries with respect to features in input space, i.e., fingerprints, which is particularly beneficial for indoor positioning where spatial locations might be separated by only a few centimeters, and
(iii) suitability for transfer learning when the size of the fingerprint database is small.
Moreover, with the advent of tiny DL, RF fingerprinting-based positioning can be performed on edge nodes.
This edge-based implementation offers several benefits, including enhanced privacy for positioning data, reduced latency, and lower system costs by eliminating the need for a central server.
In fact, \cite{avellaneda2023tinyml} has already implemented the tiny DL-based RF fingerprinting-based positioning on edge, using BLE beacons, in an indoor asset tracking scenario, achieving a $\sim88\%$ accuracy in asset location estimation.

\subsection{Localization Stemmed Upon Two-Way Ranging Features}

Two-way ranging \cite{10060797} is a method used to estimate the distance between a target device, with an unknown location, and an anchor device, with a known location, by utilizing the time-of-flight information of the ranging signals exchanged between them.
Two-way ranging paves the way for on-device localization since the distance estimation operation is conducted directly on the target device.

On-device localization has the potential to significantly enhance localization accuracy in an indoor localization scenario.
On-device localization also overcomes the large infrastructure requirements of short-range technologies and the inaccuracy issues in long-range technologies.
To enable on-device localization in an indoor localization scenario, \cite{10060797} recommends leveraging tiny DL.
Meanwhile, during the ranging process, the target device not only estimates the distance but also accumulates various ranging features as a result of processing.
For instance, if the target device is based on the ultra wide-band radio technology, then the following ranging features get collected/accumulated: distance estimates, received signal strength, and preamble symbol accumulation \cite{10060797}.
Meanwhile, in the case of long-range radio technology, the collected/accumulated ranging features are, namely, the number of successful ranging channels, calibrated distance estimates, median of the distance estimates, mean of the distance estimates, standard deviation of the distance estimates, frequency error, received signal strength, and signal-to noise ratio (SNR) \cite{10060797}.
These features are then used as input to the tiny DL-based localization algorithm, which analyzes them and returns the target device coordinates.
Relative to an indoor localization multilateration benchmark algorithm \cite{10060797}, the tiny DL-based on-device localization reduced the localization error by $\sim19\%$ and $\sim70\%$ in the case of the ultra wide-band and long-range radio technology, respectively.

\subsection{Pre-Distorter in MIMO-OFDM}

\begin{figure}[!t]
\captionsetup[subfigure]{labelformat=empty}
\centering
\begin{minipage}[t]{\columnwidth}
\includegraphics[width=\linewidth]{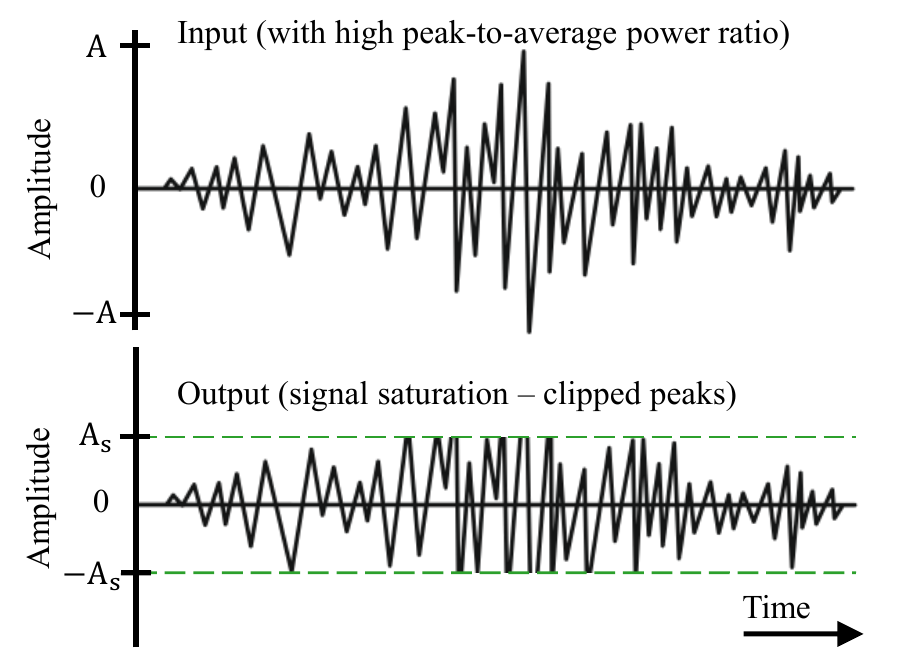}
\end{minipage}\vfill
\begin{minipage}[t]{\columnwidth}
\includegraphics[width=\linewidth]{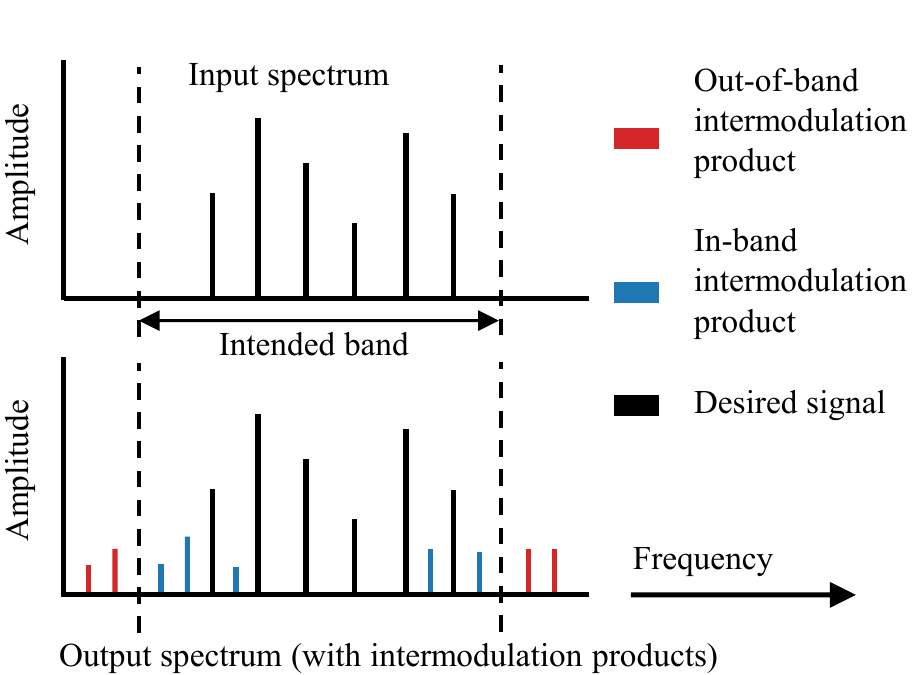}
\end{minipage}
\caption{Illustration of signal saturation and intermodulation products, caused due to high peak-to-average power ratio in the input of the nonlinear power amplifier of a MIMO-OFDM TX system.
Note that the illustration considers two domains, namely time and frequency, where the former and latter are for the signal saturation and intermodulation products case, respectively.}
\label{SS_IPs_result}
\end{figure}

A high peak-to-average power ratio in the input of the nonlinear power amplifier of a MIMO-OFDM TX system may cause signal saturation and intermodulation products, as shown in Fig.~\ref{SS_IPs_result}, which lead to distortions in the nonlinear power amplifier output \cite{10242459}.
Here, intermodulation products are in-band and out-of-band unwanted frequency components shown in Fig.~\ref{SS_IPs_result}.
This distorted signal, in turn, causes a high bit error rate on the RX side.
Introducing a pre-distorter before the nonlinear power amplifier on the TX side can resolve this issue.
The pre-distorter processes the signal with the inversed nonlinear power amplifier modeling function, thereby enhancing the linear characteristics of the nonlinear power amplifier output.
Traditional mathematical model-based pre-distorters utilize the Volterra series variants, such as memory polynomials, to approximate the aforementioned inversed nonlinear power amplifier modeling function \cite{9020606}.
In these traditional pre-distorters, the parameters of the memory polynomial are fitted based on the output of the nonlinear power amplifier.
However, this fitting process is overly sensitive to noise in the power amplifier's output, hence it leads to biased parameter configuration \cite{9020606}.
A DL-based pre-distorter is particularly effective given the non-linear nature of the power amplifier and the ability of DL to approximate a non-linear function.
However, a field-programmable gate array-based TX cannot support the high computational/memory demands of the DL-based pre-distorter.
This is where the tiny DL-based pre-distorter comes into play, as it combines low computational and memory requirements with the power of DL to approximate the inversed nonlinear power amplifier modeling function.
For instance, the size and mean execution time of the tiny DL-based pre-distorter from \cite{10242459} are $57$kB and $\sim24$s, respectively, while these grow to $263$kB and $\sim39$s in the case of a typical DL-based pre-distorter.
Besides, to achieve a bit error rate of $10^{-5}$, the SNR required in the case of the tiny DL-based pre-distorter and its DL-based counterpart are $11.22$dB and $10.54$dB, respectively \cite{10242459}.
Thus, the performance of the tiny DL-based pre-distorter is comparable to its DL-based counterpart.
On top of that, the tiny DL-based pre-distorter has a relatively smaller model size and execution time.

\subsection{Data Symbol Detection}\label{MISO_sytem_tinyML}

In a typical OFDM system, the RX explicitly estimates the channel state information (CSI) using the pilot symbols in the first block of the OFDM frame.
This CSI is then used to coherently recover/detect the data symbols in the subsequent block of the OFDM frame.
However, a DL-based approach, taking the OFDM frame as input and outputting the recovered data symbols without explicitly estimating CSI, offers relatively higher robustness to the following cases than traditional least square (LS) and minimum mean-square error (MMSE) estimation methods \cite{8052521}:
(i) fewer pilot symbols are used,
(ii) the cyclic prefix is omitted, and
(iii) the nonlinear clipping distortion effect is taken into account.

Despite its benefits, the DL-based approach is not implementable on energy-scarce RXs, making the tiny DL-based approach a promising alternative.
In fact, \cite{9745826} has already proposed a tiny DL-based approach by applying the randomized matrix approximation technique, dividing each layer of a full DL model into three sub-layers.
The division results in a forward graph/tiny DL model, whose size and inference speed are about $4.5\times$ smaller and faster, respectively, relative to its full DL-based counterpart.
On top of this, their bit error rates are similar in the cases mentioned in the previous paragraph \cite{9745826}.

\subsection{Context Sensing From Energy Harvesting Patterns}\label{contextSensing_subsection}

Consider scenarios where IoT devices, with energy harvesting capabilities, are deployed to sense their surrounding environment.
In such settings, the harvested parameters, such as energy/power or voltage/current, often inherently encode contextual information about the environment \cite{10285066, lopez2025foundations}.
For instance, thermoelectric harvesters can capture surface temperature variations through corresponding variations in harvested voltage/current \cite{lopez2025foundations}.
Additional examples can be found in \cite{8944276, 9446528}.
The paradigm of extracting environmental information by leveraging the context-inferring harvested parameters is known as context sensing.
Notably, context sensing reduces the sensing-based energy consumption on the device side, as it replaces sensors with context-inferring techniques that operate directly on harvested parameters, thus improving energy efficiency in wireless networks and reducing its deployment cost \cite{lopez2025foundations}.
The context-inferring techniques are deployed on the edge node. This ensures only high-level decisions or alerts are transmitted to the cloud. Thus, context sensing enhances privacy in wireless networks.
Meanwhile, tiny DL can be employed as a context-inferring technique, as exemplified in \cite{WANG2026121471}.
Therein, a context sensing-cum-tiny DL-based two-end framework was built to detect faults in a freight wagon bearing.
The front-end of the framework is an energy harvesting device that harvests energy by converting mechanical vibrations of a bearing into voltage responses. The changes in the physical condition of a bearing induce measurable changes in the harvested voltage responses, which serve as intrinsic indicators for physical condition, eliminating the need for additional sensors.
These voltage responses are then transmitted to the back-end, a low-power edge node. The back-end first preprocesses the received voltage responses, then feeds the processed output to a tiny DL-based classifier, which classifies between faulty and healthy physical conditions.
The two-end framework obtained $98.11\%$ and $98.15\%$ of classification accuracy and recall rate, respectively.

\subsection{Detecting Intrusions in Network Traffic} \label{intusionDetectionExisting}

Network-based intrusion detection systems (NIDSs) protect the system by detecting various types of intrusions in the network traffic.
Intrusions are generally classified into two main categories: signature- and anomaly-based \cite{9172014}.
The signature-based intrusions are detected by their known characteristic signatures.
A NIDS has to continuously update its signature database to detect any new variants of the signature-based intrusions.
Conversely, anomaly-based intrusions are detected by recognizing deviations from typical network traffic patterns.

For signature-based intrusions, ML-based NIDS can learn the general characteristics of known threats, eliminating the need for continuous signature database updates.
For anomaly-based intrusions, ML-based NIDS can continuously learn and adapt to the usual network traffic patterns, enhancing IDS's ability to detect unusual network traffic patterns over time \cite{8386762}.

\begin{figure}[!t]
\centering
\includegraphics[width=\linewidth]{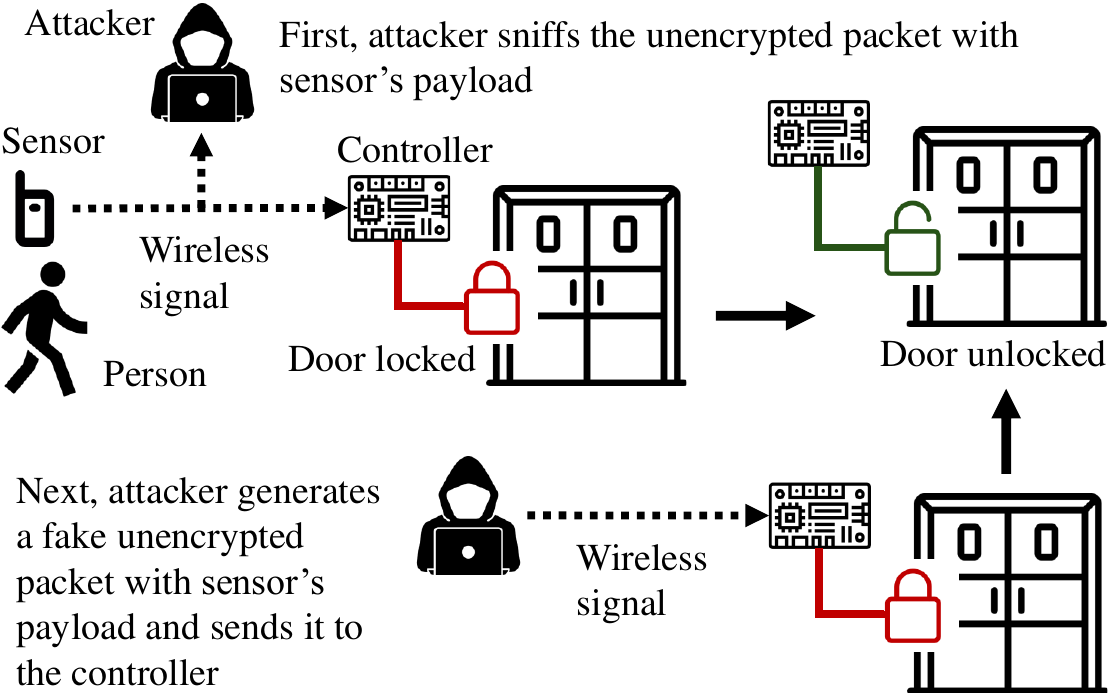}
\caption{An example of an intrusion attack on end-devices.
Upon detecting a person, the sensor sends an unencrypted packet to the controller.
The controller, after confirming the packet's payload, unlocks the door.
However, the attacker sniffs the aforementioned packet and later uses it to generate a fake packet with the sensor's payload.
The attacker then sends this fake packet to the controller, which, deceived by the legitimate-looking data, unlocks the door.}
\label{exampleIntrusionAttackfigure}
\end{figure}

Embedding an NIDS directly within the end-device/edge node may be more effective for mitigating the risks posed by intrusion attacks than deploying the NIDS solely on the central control unit \cite{10540991}.
An example of an intrusion attack targeting end-devices is illustrated in Fig.~\ref{exampleIntrusionAttackfigure}.
The advantage of having NIDS embedded in the end-device/edge node is that risk-mitigating countermeasures can be implemented immediately upon detecting an intrusion attack, even before notifying the central control unit.
These countermeasures include, first, temporarily ceasing operations of the end-device/edge node where either the unusual behaviour or anomalous injected data packets indicative of unauthorized network access have been detected, and then isolating the respective end-device/edge node from the network \cite{10540991, 10148964}.
Notably, tinyML could facilitate the implementation of NIDS within the end-device/edge node.

A tinyML-based two-layered NIDS for signature-based intrusions was proposed in \cite{fi16060200}.
In this system, the first layer of NIDS is embedded within IoT devices, and it employs the tinyML version of the extreme gradient boosting algorithm.
The first layer is responsible for classifying the network traffic as normal or an intrusion attack.
Denial of service, ransomware, scanning, and man-in-the-middle attacks are taken into account here.
The second layer of NIDS is stationed on the cloud, and it utilizes the extreme gradient boosting algorithm.
This cloud-based layer detects and classifies a broader range of intrusion attacks, including denial of service, distributed denial of service, backdoor, ransomware, cross-site scripting, man-in-the-middle, injection, brute force, and port scanning.
At the device level, the achieved NIDS's inference time and accuracy are $1.5$ms and $97.6\%$, respectively.
Meanwhile, the values of the aforementioned parameters at the cloud level are $2.24$ms and $98.3\%$, respectively.

Currently, anomaly-based intrusion detection is exclusively handled by the central control unit, as noted in \cite{10148964}.
However, as discussed in previous paragraphs, there are significant benefits of implementing NIDS directly within the end-device/edge node.
Therefore, tinyML-based NIDS, which could be embedded in the end-device, for detecting anomaly-based intrusions is a valid open research area.
Meanwhile, tiny DL algorithms from Section~\ref{ML_batteryPoweredAlgos} and \ref{ML_edgeAlgos} can be used for this purpose.

{
\setlength\arrayrulewidth{1pt}
\begin{table*}[!t]
\caption{Summary of the Deployment of TinyML-Based End-Device/Edge Inference Already Documented in the Literature\label{summary_EXISTING_table}}
\centering
\begin{tabular}{@{}p{1.45cm} p{8cm} p{8cm}@{}}
\hline
\textbf{Realm} & \textbf{Challenges} & \textbf{Solving Approaches} \\ 
\hline
Semantic data transmission & Prevent IoT devices from transmitting semantically non-relevant images to the edge node  & IoT devices use tiny DL models to extract features from images, followed by similarity measure calculation based on the semantic query sent by the edge node.
Only images that transcend the semantic query get transmitted \cite{shiraishi2023energy}  \\
\hline
Sensor linearization & Individual sensors, even of the same model, can exhibit unique manufacturing inconsistencies, temperature/environmental variations, and aging effects over time.
All of these factors uniquely influence the linearization function per sensor.
Thus, there is a need for smart on-device sensor linearization  & Tiny DL-based on-device sensor linearization approach, where the tiny DL model approximates the sensor's linearization function \cite{BOTEROVALENCIA2023e00477}  \\
\hline
Radar data & Significant target characteristic information available in radar images is lost when transformed to radar points  & Tiny DL can be used to harness the rich contextual information available in radar images to enhance the performance of radar-based perception \cite{9892925}   \\
\hline
RF fingerprinting-based positioning & Traditional DL-based RF fingerprinting-based positioning require significant computational resources and cannot be executed on an edge node  & Tiny DL-based RF fingerprinting approach for edge node-based execution.
Moreover, it offers enhanced privacy for positioning data, reduced latency, and lower system cost \cite{avellaneda2023tinyml} \\
\hline
Indoor localization & Indoor localization has technology-based limitations, such as large infrastructure requirements by short-range technologies and inaccuracy issues in long-range technologies   & Tiny DL-based on-device localization, along with two-way ranging, addresses the technology-based issues faced in indoor localization \cite{10060797}  \\
\hline
Pre-distorter & DL-based pre-distorter cannot be implemented on a field-programmable gate array-based TX in a MIMO-OFDM system  & Tiny DL-based pre-distorter can be used, since it approximates the inversed nonlinear power amplifier modeling function with low computational and memory requirements relative to its DL counterpart \cite{10242459} \\
\hline
Data symbol detection & DL-based data symbol detection approach, which recovers data symbols without explicitly estimating CSI, is not implementable on energy-scarce RXs   & Tiny DL-based data symbol detection approach can be used, since its memory requirement and inference speed are smaller and faster, respectively, relative to its DL counterpart \cite{9745826}  \\
\hline
Context sensing & Minimize the sensing-based energy consumption in wireless networks  & Replace sensors with tiny DL-based context-inferring algorithms that operate directly on harvested parameters, as they often inherently encode contextual information about the environment \cite{lopez2025foundations, WANG2026121471}  \\
\hline
IDS & Embedding the NIDS directly within the end-device/edge node   & Tiny DL can facilitate the implementation of NIDS within the end-device/edge node \cite{fi16060200} \\
\hline
\end{tabular}
\end{table*}
}

\subsection{Key Takeaways}

A summary of the challenges and respective solving approaches in already documented tinyML deployment possibilities is available in Table~\ref{summary_EXISTING_table}.
Key takeaways are as follows.
Tiny DL-based semantic communication reduces network congestion, communication overhead, and overall energy consumption by preventing the transmission of IoT data semantically non-relevant with respect to downstream tasks. This increases the IoT network's spectral efficiency and scalability.
Tiny DL-based sensor linearization reduces measurement errors and minimizes the need for repeated sensing and retransmissions, thereby improving network efficiency.
These capabilities make sensor linearization particularly valuable for resource-constrained end-devices.
In embedded radar systems, tiny DL enables real-time perception and reduces dependence on cloud processing. This makes embedded radar systems suitable for automotive and robotics-related applications.
Implementing tiny DL for on-device and edge-based indoor localization preserves user privacy, reduces response latency for users, and overcomes the technology-based issues faced in indoor localization.
In signal processing, tiny DL enables computationally efficient pre-distortion on field-programmable gate array-based TXs and CSI-free data symbol detection on energy-scarce RXs. Pre-distortion improves signal quality, while CSI-free data symbol detection reduces reliance on explicit channel estimation and increases robustness in dynamic wireless environments.
In IoT networks with energy harvesting end-devices, tiny DL enables context sensing on low-power edge nodes. This enhances privacy and energy efficiency in IoT networks.
Lastly, tiny DL strengthens network security by embedding NIDS, for both signature- and anomaly-based intrusions, directly into end-devices or edge nodes. This results in reduced communication overhead as network traffic is not continuously transmitted to the central control unit, improved privacy, and faster responses to network intrusions.

\section{Untapped Deployment Possibilities of TinyML}\label{UntappedDeplymentPossibilities}

Section~\ref{tinyMLexistingDeplymentPossibilities} covers only a handful of deployment possibilities of tinyML.
However, several potential deployment opportunities remain untapped in the existing wireless networks literature, as discussed next.

\subsection{Near-Field WPT}

\begin{figure}[!t]
\centering
\includegraphics[width=\linewidth]{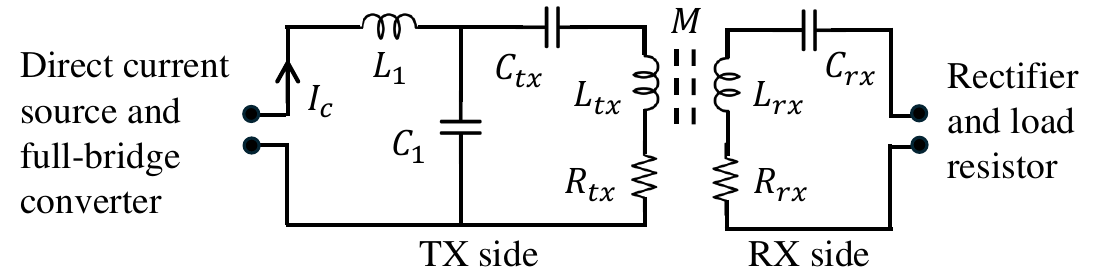}
\caption{An example of a schematic of the MRI-WPT system.
Here, $C_{tx}$ and $C_{rx}$ are compensation capacitors, $L_{1}$ is the compensation inductor, and $M$ is the mutual inductance between the TX and RX.
Moreover, the matching circuit on the TX side consists of ${\{L_{1}, C_{1}\}}$.
In the schematic, the time-varying magnetic field induced by the TX coil $L_{tx}$ gets coupled to the RX coil $L_{rx}$.
This coupling leads to the generation of voltage across the load on the RX side.
Meanwhile, $C_{tx}$ and $C_{rx}$ allow the TX and RX side to resonate at a specific frequency.}
\label{MRI_WPT_system_figure}
\end{figure}

Near-field WPT utilizes either electric or magnetic coupling to transfer power to the RX within the near-field region \cite{versloot2014optimization}.
Compared to the far-field WPT, the near-field WPT is safer for humans and has less sensitivity to directionality as power transfer here does not rely solely on a direct line-of-sight between the TX and RX \cite{versloot2014optimization}.
Our focus here is on the near-field magnetic-resonance-induction (MRI)-WPT system, where the time-varying magnetic field induced by the TX coil gets coupled to the RX coil.
This coupling leads to the generation of voltage across the load on the RX side.
The schematic of the TX and RX sides in the MRI-WPT system is shown in Fig.~\ref{MRI_WPT_system_figure}.
The compensation capacitors in Fig.~\ref{MRI_WPT_system_figure} cover the resonance segment of the MRI-WPT system, as they allow the TX and RX side to resonate at a specific frequency \cite{GRAVES2024100384, en11020352}.
Note that the schematic in Fig.~\ref{MRI_WPT_system_figure} illustrates the series-series compensation topology, while there is also a series-parallel option \cite{en11020352}.
Next, we provide a detailed description of tasks, within the MRI-WPT, for which tinyML can be used.

\subsubsection{Estimate System Characteristics With Sensorless RX}\label{estimateCharacteristicsWPT}

It is challenging to design a near-field MRI-WPT system with a freely movable RX side.
This difficulty arises from the fact that movable RX side leads to misalignment between the TX and RX coils, which results in fluctuations in the characteristics of a near-field MRI-WPT system \cite{11015570}.
These characteristics include the load resistance on the RX side, the power delivered to the load resistor on the RX side, and the coupling coefficient between the TX and RX sides.

The load resistance and coupling coefficient impact the power delivered to the load and WPT-efficiency, respectively \cite{s24020501}.
Therefore, it is crucial for the TX side to be aware of the MRI-WPT system characteristics to optimize the system performance.
Note that the information about the RX side's current/voltage measurements is required to compute the MRI-WPT system characteristics \cite{9756526}.
To transfer this information, the RX side must be equipped with sensors and communication circuit, increasing manufacturing costs and the standby power consumption on the RX side.
Hence, a sensorless-cum-movable RX, wherein the TX estimates the MRI-WPT system characteristics using only TX current/voltage measurements is an appealing approach.
In such a case, the input current $I_c$ of the front-end compensation inductor \cite{9756526} and the capacitor voltage of the matching circuit on the TX side \cite{s24020501}, i.e., the voltage across $C_1$, can be put to use to estimate the MRI-WPT system characteristics.
Unfortunately, when either there is weak coupling among the TX and RX sides or the load resistance is high, equation-based estimation approaches result in high estimation errors \cite{9756526}.
This issue paves the way for the utilization of tiny DL-based estimation approach, which can take into account the non-linearities, such as higher-order harmonics in the input current of the front-end compensation inductor, to address the aforesaid high estimation error scenarios.
In particular, tiny DL models, built upon the frameworks from Section~\ref{ML_edgeAlgos}, can be used for the aforementioned system characteristics estimation.
The following are examples of low-power TXs in the MRI-WPT systems: wireless chargers for wearable IoT devices/consumer electronics (mobile phones)/biomedical implants, low-power drone recharging platform, etc.

\subsubsection{Activate/Deactivate a Given TX-Coil in a Multi-TX Systems}

In a multi-TX near-field MRI-WPT system featuring low-power TXs and movable sensorless RX, tiny DL can be employed to determine whether to turn on/off a given TX coil at any given time.
The tiny DL-based estimator would analyze the input current of the front-end compensation inductor at the TX side and output the estimated power delivered to the load resistor at the movable RX side.
Meanwhile, tiny DL models, devised using the frameworks from Section~\ref{ML_edgeAlgos}, can be used for the aforementioned power estimation.
Note that this estimated power is directly proportional to the WPT efficiency.
Next, the decision about whether to turn on/off a given TX coil would be taken based on the expected WPT efficiency \cite{9756526}.

\subsubsection{Estimate Compensation Capacitances of the Near-Field MRI-WPT System With Ferrite Shields}

The schematic of the near-field MRI-WPT system with ferrite shields is similar to the one in Fig.~\ref{MRI_WPT_system_figure}.
A key challenge here is the absence of analytical models to compute the self-inductance of coils present on the TX and RX sides, the parasitic resistances of these coils, and the mutual inductance between these coils.
As a result, the TX and RX side's compensation capacitances that tune with the resonant frequency of the MRI-WPT system and maximize the power delivered to the RX's load resistor cannot be computed analytically.

In \cite{10400566}, the authors suggest using a field solver \cite{field_Solver} to determine compensation capacitances given the resonant frequency of the MRI-WPT system and the distance between the TX and RX coils.
However, this approach is limited by its high computation time, rendering it impractical for real-time usage.
This motivates the use of a tiny DL-based estimator, which can model the nonlinearity between design parameters and compensation capacitances.
Meanwhile, tiny DL models, built upon the frameworks from Section~\ref{ML_edgeAlgos}, can be used for the aforementioned compensation capacitance estimation.
It is essential to implement the tiny DL-based estimator on both the TX and RX side, since compensation capacitors are present on both sides.
Once compensation capacitances are estimated, the corresponding compensation capacitors can be tuned accordingly.

\subsubsection{Maintain Constant Power at the Load Resistor}

Usually, there are variations in the power delivered from the same TX to RX load resistors with different compensation topologies.
Additionally, the power delivered to the load resistor in the near-field MRI-WPT system tends to fluctuate over time due to changes in either load resistance or mutual inductance between the TX and RX coils or both.
Note that obtaining real-time values for load resistance and mutual inductance is challenging because the TX and RX sides are physically separated.
Furthermore, the power delivered to the load resistor is also influenced by the capacitance of the compensation capacitor of the TX side \cite{10521777}.
Thus, tuning the TX side's compensation capacitance can be helpful to ensure that constant power is delivered to the load resistor, regardless of changes in RX topology, load resistance, or mutual inductance \cite{10521777}.
Since compensation capacitance cannot be computed analytically, one may leverage the tiny DRL, e.g., using approaches from Section~\ref{ML_edgeAlgos}.
In such a case, the \textit{action} would be the estimated compensation capacitance, the \textit{state} would consist of the voltage across the compensation capacitor and the current flowing from the TX coil, and the \textit{reward} would be a function of the estimated power delivered to the load resistor.

\subsubsection{ULD in Unmanned Aircraft System}

In the case of unmanned aircraft systems (UASs), one way to increase their flight time is to create near-field MRI-WPT-based recharging platforms where they can land to recharge their batteries.
The near-field MRI-WPT systems from \cite{GRAVES2024100384, en11020352} can be used for this purpose.
However, coupling the TX side of the MRI-WPT system with an unintended object can induce eddy currents in that object.
These unwanted currents increase the temperature of that object, leading to a safety hazard \cite{GRAVES2024100384}.
The problem of identifying an unintended object and restricting it from coupling with the TX side is known as ULD.

\begin{figure}[!t]
\centering
\includegraphics[width=\linewidth]{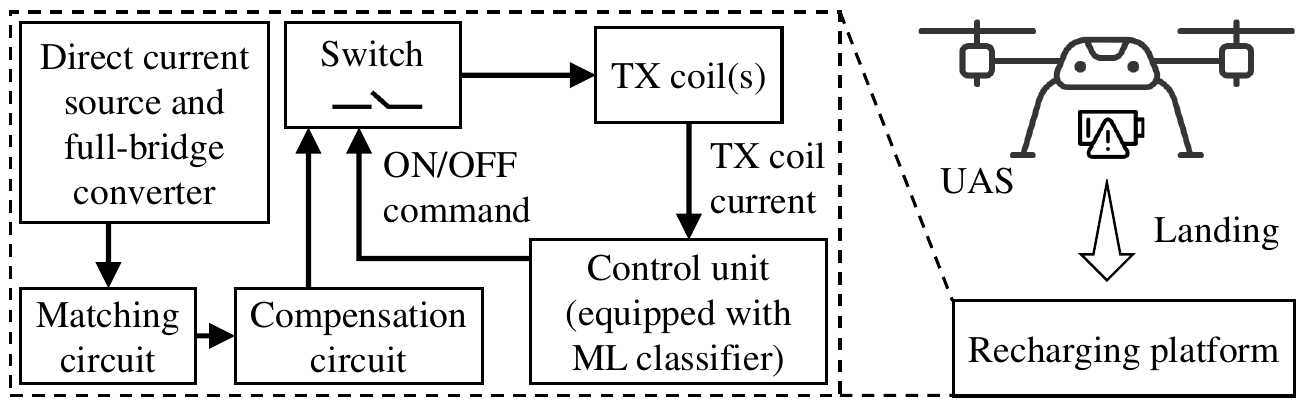}
\caption{The schematic of a near-field MRI-WPT-based recharging platform capable of solving the ULD problem in UAS.
Here, UAS with a low-battery status is landing on the recharging platform to recharge itself.
Initially, the switch is ON, i.e., there is a connection between the compensation circuit and the TX coil.
Next, the control unit analyzes the TX coil's current waveform characteristics and provides the ON/OFF command to the switch to carry-on/stop the WPT to UAS.}
\label{ULD_system_figure}
\end{figure}

In UAS, one potential solution to the ULD problem is to establish a communication link between the TX and RX sides.
This communication link would later function as a feedback loop, ensuring that only the intended RX load receives power.
However, in systems, like UAS, where maintaining a constant/stable power density is considered as a crucial design feature, the signal exchange needed to establish a communication link can reduce the power density on the RX side.
Interestingly, the current waveform characteristics of the TX side coil are unique to the RX side load \cite{GRAVES2024100384}.
This implies that a change in the RX side load alters the TX coil's current waveform characteristics.
Therefore, by analyzing the TX coil's current waveform characteristics using a tiny DL-based intended/unintended RX load classifier at the TX side, the ULD problem could be solved without even establishing a communication link between the TX and RX sides.
The schematic of a near-field MRI-WPT-based recharging platform capable of solving the ULD problem is shown in Fig.~\ref{ULD_system_figure}.
Again, tiny DL models, devised using the frameworks from Section~\ref{ML_edgeAlgos}, can be used as the aforementioned classifier.

\subsection{Backscatter-Assisted Wireless Sensor Networks}

Backscattering is a type of passive communication technology, where nodes modulate and reflect the existing RF signal to transmit their data rather than generating their own RF signals \cite{lopez2024zero}.
Backscattering systems consist of three components: carrier emitter, backscatter node, and reader.
The carrier emitter emits the carrier RF signals, while the backscatter node modulates and reflects its incident carrier RF signal towards the reader \cite{zargari2024deep}.
Here, the reflected signal is known as the backscattered signal.
Meanwhile, there are three types of backscattering: monostatic, bistatic, and ambient \cite{lopez2024zero, zargari2024deep}.
In monostatic, the reader also acts as the carrier emitter and demands duplexing capabilities.
In bistatic, the reader and carrier emitter are not co-located.
In ambient, there is no carrier emitter, and backscatter nodes modulate plus reflect the incident ambient RF signals towards the reader.

Consider the backscatter-assisted WSN (BA-WSN), where a batteryless sensor node can perform either one of the following three functions at any time step \cite{8647737}:
(i) perform active RF transmission of its own data,
(ii) harvest energy from incident RF signals transmitted by other sensor nodes to support its own future RF data transmissions, and
(iii) harvest a portion of the energy of the incident RF data signal, transmitted by other sensor nodes, and then utilize the remaining portion to relay that RF data signal to the sink by backscattering it, without modulating/demodulating.
The sensor nodes that act as relays can adjust the portion of energy to be utilized to backscatter by tuning their reflection coefficients.
In addition, the reflection coefficients may be tuned to allow the backscattered signals to combine constructively at the sink by controlling the multipath fading, resulting in enhancement of the received RF data signal strength.
Improved signal strength, in turn, boosts the sum rate at the sink \cite{8647737, zargari2024deep}.
In such a setup, sensor nodes may not share any kind of information among themselves.
Thus, tiny RL, e.g., using approaches from Section~\ref{ML_batterylessAlgos}, with a discrete action space can be used to devise the reflection coefficient update policy at each sensor node.
Here, the tiny RL algorithm's \textit{action} would be the selection of the reflection coefficient, while the \textit{state} vector could take into account the real/imaginary parts of the backscatter channel vector \cite{zargari2024deep}, incident signal power, and \textit{action}/\textit{reward} from the previous relay instance.
The \textit{reward} could capture the spectral efficiency and penalties for \textit{actions} leading to too low energy harvesting.
In this setup, the sink may broadcast its received signal's data rate, a proxy for the spectral efficiency, to sensor nodes.
Finally, since the backscattered signal is vulnerable to eavesdropping attack \cite{11087490}, appropriate defense measures, such as backscatter masking, to counter eavesdropping must be taken into account. Moreover, for scenarios where sensor nodes are densely co-located, the reflection coefficient update method must also take into account the near-field sensor-to-sensor coupling issue \cite{11087490}.

\begin{figure}[!t]
\centering
\includegraphics[width=\linewidth]{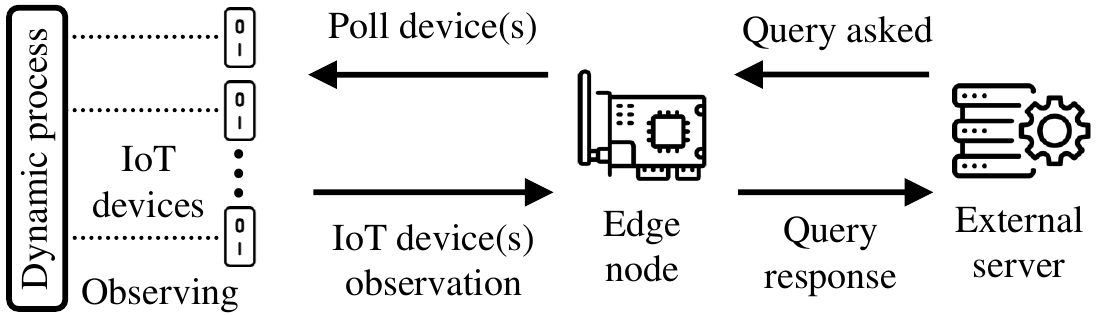}
\caption{Illustration of a GoC system.
The external server asks queries about the dynamic process state, observed by IoT devices, to the edge node.
Next, the edge node polls a subset of devices and estimates the complete state, with the help of the received information and a state estimator.
Finally, the edge node responds to the query by using the estimated state.}
\label{GoC_system_figure}
\end{figure}

\subsection{Goal-Oriented Communication}

Goal-oriented communication (GoC) is a paradigm, which aims at identifying and transmitting only the data relevant to the task taken in hand by the RX \cite{10884945}.
This makes GoC befitting for scenarios with either communication resource scarcity, or IoT devices/edge node as TXs/RX, or both.
The following are some use cases of GoC, wherein tinyML can be utilized.

\subsubsection{Goal-Oriented Dynamic System Monitoring}\label{GoC_system_Monitoring}

Consider the GoC system from \cite{raghuwanshi2024goal}, as shown in Fig.~\ref{GoC_system_figure}, where IoT devices observe the dynamic process's state, while external servers ask queries about the aforementioned state to the edge node, whose goal is to minimize the mean square error in the query response $(\textrm{MSE}_{q})$.
Moreover, the edge node has no information about the server's query process.
Thus, the goal-oriented device scheduling problem in this system can be modeled as a partially observable Markov decision process (MDP), wherein the edge node acts as the \textit{agent} with \textit{action space} $\{0,1,\cdots,N\}$.
Action $n\in\{1,\cdots,N\}$ refers to polling device $n$, while action $n=0$ refers to withdraw from polling.
The agent's \textit{state} would be a function of the prior estimates and query-related parameters such as the time elapsed since the last query from external servers \cite{raghuwanshi2024goal}, while its \textit{reward} would be a function of $\textrm{MSE}_{q}$.
Note that the edge node is equipped with a state estimator to estimate the prior estimates, obtained before device polling, of the dynamic process's state.

\begin{figure}[!t]
\captionsetup[subfigure]{labelformat=empty}
\centering
\begin{minipage}[t]{0.5\columnwidth}
\includegraphics[width=\linewidth]{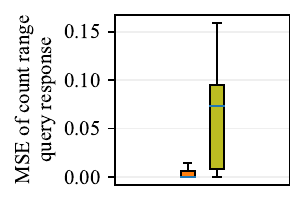}
\end{minipage}\hfill
\begin{minipage}[t]{0.5\columnwidth}
\includegraphics[width=\linewidth]{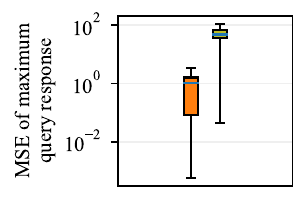}
\end{minipage}\vfill
\begin{minipage}[t]{\columnwidth}
\includegraphics[width=\linewidth]{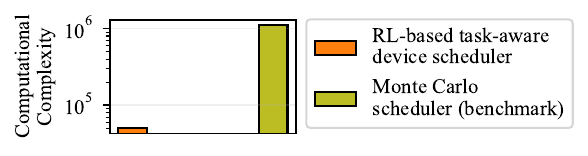}
\end{minipage}
\caption{Box plots for $\textrm{MSE}_{q}$ accumulated during scheduler run and the computational complexity of a single scheduling operation.
Note that we have quantified the computational complexity of schedulers in terms of the number of arithmetic operations they performs.}
\label{GoC_result}
\end{figure}

The edge node can exploit RL to discover the device scheduling policy for the aforementioned partially observable MDP.
To test RL's performance, let us consider the nonlinear dynamic process, count range query, and maximum component query from \cite{raghuwanshi2024goal}.
Fig.~\ref{GoC_result} shows that relative to the Monte Carlo benchmark scheduler, which polls IoT devices only when a query arrives, the RL-based scheduler has minimized $\textrm{MSE}_{q}$ for both query types and the computational complexity of a single scheduling operation.
However, the RL-based scheduler exhibits a peak RAM usage of $\sim552$ megabytes, which is considerably high from the standpoint of a low-power edge node.
Thus, we need a tiny RL, e.g., approaches from Section~\ref{ML_edgeAlgos}, to empower the low-power edge node with the RL-based scheduler, and procure the performance gain shown in Fig.~\ref{GoC_result}.

\subsubsection{Goal-Oriented Control System}

\begin{figure}[!t]
\centering
\includegraphics[width=\linewidth]{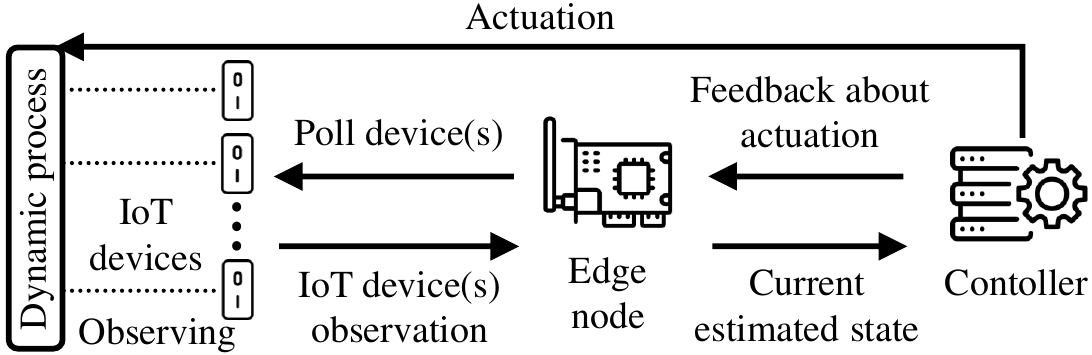}
\caption{Illustration of a goal-oriented control system.
At each time step, the controller asks about the dynamic process state, observed by IoT devices, to the low-power edge node.
Next, the edge node polls a subset of devices, estimates the complete state using a state estimator, and responds to the controller.
Finally, the controller sends feedback about the recently performed actuation to the edge node.}
\label{GoC_control_figure}
\end{figure}

Consider the GoC system shown in Fig.~\ref{GoC_control_figure}, where IoT devices observe the dynamic process's state.
At each time step, the low-power edge node provides information about the dynamic process's state to the controller, which performs the actuation, and in return receives the actuation feedback.
In this system, the goal is to minimize the mean square error in the actuation operation $(\textrm{MSE}_{a})$, while the goal-oriented device scheduling problem can be modeled as a MDP, wherein the edge node acts as the \textit{agent} with \textit{action space} $\{0,1,\cdots,N\}$ from Section~\ref{GoC_system_Monitoring}.
Considering the edge node is equipped with a state estimator.
Then, the agent's \textit{state} would be a function of the state estimator's prior estimates, while its \textit{reward} would be a function of $\textrm{MSE}_{a}$.
Meanwhile, the edge node can exploit tiny RL, e.g., using approaches from Section~\ref{ML_edgeAlgos}, to discover the device scheduling policy for the aforementioned MDP.

\subsubsection{Goal-Oriented Push/Pull Coexistence}

Consider a system involving both push-based and pull-based communication.
The pull-based communication follows the framework from Section~\ref{GoC_system_Monitoring}, while the decision to perform push-based communication is taken autonomously by IoT devices.
In addition to responding to data requirements from the edge node, an IoT device also heeds for anomalies in its observed data.
Once an anomaly is detected, the IoT device sends that anomalous observation to the edge node through push-based communication.
Detection of anomalous observations might be missed in the case of pull-based communication, however, push/pull coexistence resolves this issue.
Meanwhile, in the goal-oriented push/pull coexistence system, tiny RL can be used to discover the device scheduling policy for pull-based communication, as mentioned in Section~\ref{GoC_system_Monitoring}.
Furthermore, tiny DL models, built upon the frameworks from Section~\ref{ML_batteryPoweredAlgos}, can be used for the purpose of anomaly detection in the observed data.

\subsection{Local Wake-Up Strategy} \label{wakeupStrategy}

Consider the alarm scenario where batteryless IoT devices are deployed to sense the occurrence of alarm events in a geographical area, as in \cite{ruiz2024intelligent}.
Once an alarm event is sensed, the devices send the event information to the edge node.
To prolong the time for which the geographical area is being sensed by batteryless devices, the duty cycle of these devices must be managed intelligently by an on-device wake-up radio (WuR) module, which enables them to operate in the following states \cite{ruiz2024intelligent}:
idle, active, transmission, and sleep.
Such duty cycle management would need to minimize the probability of missing alarm events.

\begin{figure}[!t]
\centering
\includegraphics[width=\linewidth]{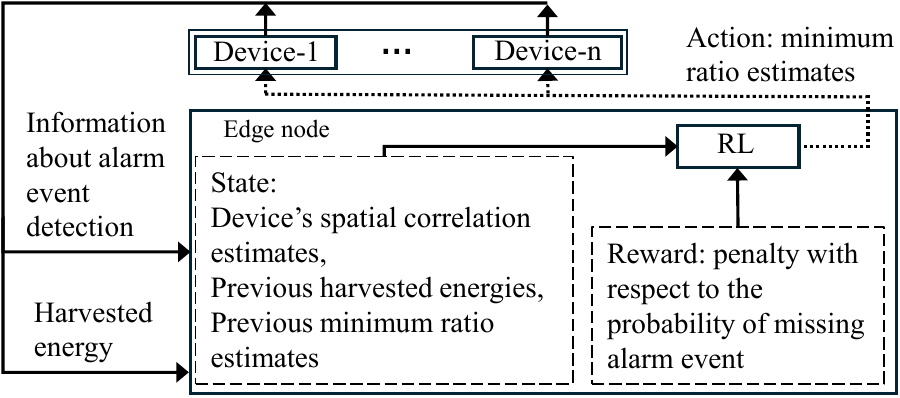}
\caption{Illustration of the modus operandi to minimize the probability of missing alarm events.
First, the edge node provides the state vector as input to its RL algorithm, which then estimates the desired minimum value of the ratio of the device's energy level for the upcoming time interval of fixed duration.
Next, the aforementioned minimum ratio estimates are sent to the WuR of each device, which utilizes the received information to devise the duty cycle for the aforementioned time interval.
Later, the edge node receives information about the harvested energy and alarm event detection, during the aforementioned time interval, from devices.
Lastly, a reward is provided to the RL algorithm by taking into account the received information at the edge node.}
\label{WuR_system_figure}
\end{figure}

The aforementioned problem can be solved with the following modus operandi, illustrated in Fig.~\ref{WuR_system_figure}.
First, for each device, the edge node estimates the desired minimum value for the ratio of the device's energy level and its maximum energy storage capacity.
Note that these ratios are estimated with the goal to minimize the probability of missing alarm events in the upcoming time interval of fixed duration.
Meanwhile, the edge node can utilize tiny RL, e.g., using approaches from Section~\ref{ML_batterylessAlgos} supporting the continuous state/action space, for this purpose.
Here, tiny RL's \textit{action} would be the aforementioned minimum ratio estimates, while its \textit{state} vector would take into account the spatial correlation among devices as well as the harvested energies and the aforementioned minimum ratio estimates from previous time intervals.
Besides, its \textit{reward} would take into account penalties with respect to the probability of missing alarm events, to prevent \textit{action} that results in power failure on device(s).
Next, the edge node provides the aforementioned minimum ratio estimates to the WuR of each device, which utilizes the following parameters along with the received estimates to devise the duty cycle for the aforementioned upcoming time interval:
available energy, harvesting power, power required by the sensing task, and maximum energy storage capacity.
Meanwhile, a device can utilize the information acquisition methods from \cite{lopez2025foundations} to infer the first three parameters.

\subsection{Underwater Visible Light Channel Estimation}

Consider the scenario of an underwater visible light communication (UVLC) between an underwater fixed platform and a submersible edge node.
In such system, accurate underwater channel estimation is crucial as it improves both the post- and pre-equalization of the signal in end-to-end UVLC, as emphasized in \cite{Cai_23}.
This paves the way for the tiny DL-based channel estimator, which would effectively estimate the underwater channel by estimating the following three key distortions:
linear distortion from inter-symbol interference \cite{Cai_23},
non-linear distortion from the optoelectronic devices such as photodiode and light-emitting diode \cite{Cai_23}, and
quadratic distortion from signal-to-signal beat interference \cite{Cai_23}.
Meanwhile, frameworks from Section~\ref{ML_edgeAlgos} can be utilized to devise the tiny DL model for channel estimation in UVLC.

Note that the low-pass/band-pass filtering from the optoelectronic devices and amplifiers in the UVLC system leads to the inclusion of a strong low-frequency noise in the received signal at the edge node.
This reduces the received signal's SNR and channel response estimation accuracy \cite{Li_22}.
To combat it, one may design a tiny DL-based channel estimator for a baseband frequency range.
Next, up-convert the signal at the TX to a frequency range whose lower limit is higher than the upper edge of the frequency region dominated by the low-frequency noise and upper limit is lower than the cut-off frequency of the UVLC system.
Lastly, down-convert the received signal to the baseband frequency range of the tiny DL-based channel estimator \cite{Li_22}.

\subsection{Desired Signal Reception and Received Data Deciphering}

In wireless communication systems, transmitted signals propagate through complex communication channels where they suffer attenuation, fading, noise, interference, etc.
As a result, the RX must employ sophisticated techniques for \textit{desired signal reception} and \textit{received data deciphering}, which are required to recover the desired information accurately and reliably from received signals.
Next, we discuss in detail the tasks, within the scope of desired signal reception and data deciphering, that can benefit from tinyML.

\subsubsection{Automatic Modulation Classification}

Consider a scenario where the end-devices straight away proceed with the data transmission.
Here, the edge node can resort to the AMC method to identify the modulation type and decipher the received data.
Although optimal classification results are guaranteed from the traditional likelihood-based AMC methods, they are poised by high computational complexity, sensitivity to unknown channel conditions, and demands regarding the prior knowledge of received signal characteristics such as SNR and frequency offsets \cite{9220797, 10949587}.
Meanwhile, traditional feature-based AMC methods have low computational complexity, but are poised by suboptimal classification results, large computation time, inability to handle signals with complex modulation, carrying out the classification task based on handcrafted features, while those handcrafted features might not be the best for the modulation type available in the received signal \cite{10949587, 9220797}.
In contrast, tiny DL-based AMC can overcome the shortcomings of the aforementioned traditional methods by autonomously extracting the desired features, from the received signals, for the classification task.
The frameworks from Section~\ref{ML_edgeAlgos} can be used to devise the tiny DL model for AMC.
Moreover, utilizing the tiny DL model reduces the bandwidth required to perform the model deployment/update task on the edge node.

\subsubsection{Predict Automatic Gain Control Index Range of RX}

In wireless communication, interference with high signal strength saturates RXs, disrupting the reception of their desired signals.
To resolve this saturation problem, the RX must constantly monitor its received signal strength and correspondingly adjust its automatic gain control (AGC) \cite{joly2024algorithm} index.
Typically, the RX adjusts its AGC index before the start of its reception period.
However, the interference signal strength may change during the RX's reception period, causing the RX to saturate, which results in data loss.
To mitigate this, it is optimal to derive a range for the AGC index to be used during the next reception period.
This would help the RX remain resilient to interference changes during the reception period.
In this context, tiny DL models, devised using the frameworks available in Section~\ref{ML_batteryPoweredAlgos} and \ref{ML_edgeAlgos}, can be employed to predict the appropriate AGC index range by analyzing RSSI, SNR, access address detected (AAD), and cyclic redundancy check metrics derived from the previously received signals.
Note that the \textit{access address} is a field in the received packet header used to identify a specific wireless connection, while AAD is a binary metric. Here, AAD~$=1$ when the received access address matches with RX's configured access address, otherwise AAD~$=0$ \cite{10556113}.

\subsection{Cognitive Radio Network}

A cognitive radio network is a network of devices, named secondary users (SUs), that dynamically adapt their communication parameters, such as transmission power, subcarrier frequency, etc, by sensing and learning from the radio environment to efficiently utilize white spaces in available spectrum while minimizing interference to the spectrum's licensed users, named primary users (PUs) \cite{muzaffar2024review}.
Next, we provide a detailed description of tasks, within the scope of the cognitive radio network, for which tinyML can be used.

\subsubsection{Cooperative Spectrum Sensing}

Consider a cognitive radio network with full-duplex SUs sensing the spectrum licensed to PUs.
In such network, SUs need a spectrum sensing strategy to improve their throughput and minimize interference to PUs.
CSS \cite{9127944} is a type of spectrum sensing strategy, where the SUs send their locally sensed information about the spectrum to the edge node, which later fuses its received information and takes the final sensing decision.
Due to the full-duplex nature, the SUs encounter both self-interference and co-channel interference.
A tiny DL-based CSS here would allow the edge node to learn temporal correlation dynamics from the SU's locally sensed information, including the activity pattern of the PUs \cite{9127944} and self-interference suppression coefficient \cite{9127944}, defined as the ratio of residual self-interference power and transmit power, of the SUs.
The frameworks from Section~\ref{ML_edgeAlgos} can be used to devise the tiny DL model for CSS.

Note that appropriate defense measures to counter learning-based attacks, such as the learning-evaluation-beating attack \cite{9220838}, must be taken into account in the case of tiny DL-based CSS as these attacks try to flip the edge node's sensing decision by sending false data in a way such that there is a minimum manipulation in the data pattern.

\subsubsection{Resource Allocation in a Device-to-Device Underlay Network}

Consider a device-to-device (D2D) underlay network, where the resources, such as subcarriers and power, for the cellular users (CUs) are shared by D2D pairs.
Here, the D2D pair consists of two neighboring devices in the underlay network.
Through D2D communications, the devices involved in a D2D pair can communicate directly with each other, bypassing the need for a central entity.
However, CUs and D2D pairs may experience mutual interference from neighboring D2D pairs utilizing the same subcarriers \cite{kaiUnderlayRA2023}.
This necessitates a D2D resource allocation strategy/policy to be devised at each D2D pair in a distributed manner.
Hence, a multi-agent RL approach, where each D2D pair acts as an RL agent is appealing here.

Tiny RL, e.g., using approaches from Section~\ref{ML_batteryPoweredAlgos} and \ref{ML_edgeAlgos}, can be used to devise the resource allocation policy.
Here, the tiny RL's \textit{action} could be, for instance, the selection of both the subcarrier and transmission power, while the \textit{state} vector for a D2D pair could take into account the \textit{action}/\textit{reward} from the previous transmission instance and channel vectors for the channel between the respective D2D pair-to-CU, respective D2D pair-to-adjacent D2D pair, and devices involved in the respective D2D pair.
Furthermore, the \textit{reward} could take into account the sum data rate of all D2D pairs and penalty for the \textit{action} that leads to significant data rate losses.
For this, the agents may send the information about their respective data rate to the base station catering CUs, which then computes the sum data rate and broadcasts it to agents.

\subsection{Specific Emitter Identification}\label{SEI_subsection}

Having SEI executed at edge nodes is preferred to minimize the communication overhead in IoT networks \cite{9372779}.
Unlike cryptography-based emitter identification, which is vulnerable to emitter spoofing, RF-based SEI relies on RF features that are inherent traits of RF hardware circuits.
Thus, these RF features are inherently resistant to counterfeiting.
It is worth mentioning that both RF fingerprinting-based positioning, from Section~\ref{RF_fingerprinting_positioning}, and RF-based SEI take into account RF features.
However, RF features in RF fingerprinting-based positioning are dependent on the spatial properties of the environment, while those in RF-based SEI are dependent on physical hardware traits.
Next, the RF-based SEI decomposes the SEI operation into the following two parts, namely feature embedding, which maps the RF signal sample space to the RF feature space, and emitter classification \cite{9846906}.
The feature embedding can be performed using tiny DL algorithms, built upon the frameworks from Section~\ref{ML_edgeAlgos}.

Note that variation of RF features with respect to the emitter temperature must be factored in while designing a tiny DL model for SEI \cite{s25072293, info14090479}.
The reason is that all emitters undergo a warm-up period, where their temperature first increases and later stabilizes, once they start transmitting.
Notably, the aforementioned RF features may change with emitter temperature during the warm-up period \cite{info14090479}.

\subsection{BLE-based Navigation and Proximity Detection}

BLE is widely used for navigation in indoor environments, such as hospitals, shopping malls, factories, university buildings, etc, where signals from the global positioning system are unreliable \cite{filus2022cost}.
In BLE-based indoor navigation, mobile devices estimate their position by measuring RSSI from nearby BLE beacons.
Note that RSSI represents the power level of a received radio signal and is commonly used to identify proximity among devices \cite{10550904}.
The following provides a comprehensive description of tasks associated with BLE-based navigation and proximity detection, where tinyML can be used.

\subsubsection{RSSI-Based Transmission Power Management on BLE Beacons}

Consider an indoor navigation system where MCU-based BLE beacons transmit periodic messages to movable user devices, enabling them to detect their proximity to a beacon.
In such a navigation system, tuning the transmission power of beacons is an effective strategy to mitigate interference, thereby enhancing the navigation system performance, as stated in \cite{9209253}.
This tuning is feasible because beacons can compute possible interference in their respective transmissions by calculating the RSSI of each other's transmissions.
Here, tiny RL offers a promising solution for continuously tuning the transmission power of the beacon directly on the beacon.
In such a case, received signals could act as the \textit{state}, transmission power of the beacon as the \textit{action}, while the \textit{reward} could be a function of resulting RSSI after performing the \textit{action}.
Tiny RL, e.g., using approaches from Section~\ref{ML_batteryPoweredAlgos}, can be used for the aforementioned power management task.

\begin{figure}[!t]
\centering
\includegraphics[width=\linewidth]{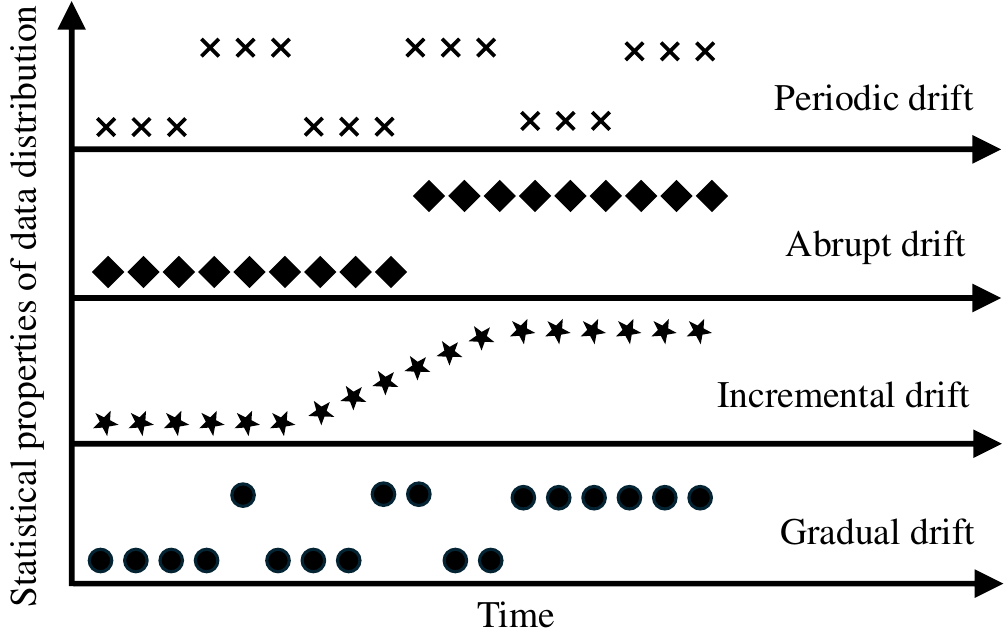}
\caption{Different types of concept drifts that occur due to the variation in the statistical properties of the data.}
\label{conceptDriftTypesfigure}
\end{figure}

\subsubsection{Proximity Detection Stemmed Upon RSSI-Based Ranging}

The distance between devices may be estimated through RSSI-based ranging.
The classical RSSI-based ranging method relies solely on spatial features and disregards the RSSI fluctuations in both the time and frequency domain \cite{9371718}.
Moreover, when classical RSSI-based ranging is applied among mobile devices, the relative motion between devices introduces additional fading to the measured RSSI, leading to reduced accuracy or confidence \cite{9371718}.
This is a clear motivation for incorporating ML into RSSI-based ranging.
Indeed, ML may explore the distance estimation by exploring a broader range of RSSI features, including \cite{9371718}:
\begin{itemize}
  \item Time-domain features: fade duration, level crossing rate, $50\%$ coherence time, Rayleigh parameter.
  \item Frequency-domain features: energy of Doppler spectrum, Laplacian best fit, root mean square (RMS) Doppler spread.
  \item Statistical features: mean, peak-to-peak change, standard deviation, interquartile range, skewness, kurtosis.
\end{itemize}
Furthermore, tiny DL models, devised using the frameworks from Section~\ref{ML_batteryPoweredAlgos}, are a viable option to implement RSSI-based ranging on a mobile MCU-based receiving end.

{
\setlength\arrayrulewidth{1pt}
\begin{table*}[!t]
\caption{Summary of the Untapped Deployment Possibility of TinyML-Based End-Device/Edge Inference\label{summary_UNtapped_table}}
\centering
\begin{tabular}{@{}p{1.45cm} p{8cm} p{8cm}@{}}
\hline
\textbf{Realm} & \textbf{Challenges} & \textbf{Research Direction} \\ 
\hline
Near-field MRI-WPT & Analytical approaches, to estimate the characteristics of the MRI-WPT system with a movable sensorless RX, result in high estimation errors when either there is weak coupling among the TX and RX sides or the load resistance is high \cite{9756526}  & Use tiny DL-based estimator, which takes into account the higher-order harmonics in the input current of the front-end compensation inductor to address high estimation error issue  \\  \hhline{~--} 
 & Turn on/off a given TX coil in a multi-TX MRI-WPT system with a movable sensorless RX \cite{9756526} & Use tiny DL-based estimator to estimate the power delivered to the RX's load resistor by analyzing the input current of the front-end compensation inductor at the TX side.
 The decision to turn on/off a given TX coil could be based on the estimated power \\  \hhline{~--}
 & Absence of analytical models to estimate compensation capacitances, at the TX and RX sides, of the MRI-WPT system with Ferrite shields.
 As a result, it is not possible to tune compensation capacitances that maximize the power delivered to the load resistor at the desired resonant frequency \cite{10400566} & Tiny DL-based estimator to model the nonlinearity between design parameters and compensation capacitances  \\  \hhline{~--}
 & Maintain constant power at the load resistor in the MRI-WPT system, regardless of changes in RX topology, load resistance, or mutual inductance \cite{10521777}  & Tiny DRL to optimize the TX side's compensation capacitance by taking into account the voltage across the corresponding TX capacitor and the current flowing from the TX coil with the aim to maintain constant power at the load resistor  \\  \hhline{~--}
 & ULD in the MRI-WPT system without establishing a communication link between the TX and RX sides \cite{GRAVES2024100384}  & Analyze the TX coil’s current waveform characteristics using a tiny DL-based classifier at the TX side to classify the intended/unintended RX side load  \\
\hline
Backscattering & Update/tune the reflection coefficients of sensor nodes in the BA-WSN to ameliorate the spectral efficiency at the sink \cite{zargari2024deep}  & Use tiny RL to distributively update/tune the reflection coefficient at each sensor node  \\
\hline
GoC & Accurately respond to queries, from external servers, about the dynamic process state \cite{raghuwanshi2024goal}  & Devise a tiny RL-based device scheduler to gather such state data that would lead to the minimization of the query response error  \\  \hhline{~--}
 & Detect anomalies in the dynamic process state along with accurately respond to queries  & Tiny DL for anomaly detection and tiny RL-based device scheduler to gather such state data that minimizes error in query responses \\  \hhline{~--}
 & Accurately perform the actuation to stop the dynamic process from going into an unstable state  & Devise a tiny RL-based device scheduler to gather such state data that would lead to the minimization of the actuation error  \\
\hline
WuR & Prolong the time for which the geographical area is being sensed, for the occurrence of alarm events, by batteryless devices \cite{ruiz2024intelligent}  & Let the edge node use tiny RL to estimate the minimum ratio estimates, from Section~\ref{wakeupStrategy}, that minimize the probability of missing alarm events.
Next, share these estimates with each device's WuR, which it uses to devise its duty cycle  \\
\hline
Channel estimation & Low-pass/band-pass filtering in the UVLC system leads to a strong low-frequency noise in the received signal. This reduces the received signal's SNR and channel response estimation accuracy \cite{Li_22}  & Devise a tiny DL-based channel estimator for a baseband frequency range. Then, up-convert the signal at the TX. Lastly, down-convert the received signal to the aforementioned baseband frequency range \\
\hline
AMC & AMC method free from the limitations of traditional likelihood-based and feature-based AMC methods \cite{9220797, 10949587}  & Use tiny DL-based AMC method. Unlike traditional methods, it autonomously extracts the desired features from the received signals, does not demand prior knowledge of received signal characteristics, and has low computational complexity   \\
\hline
Cognitive radio network & CSS in a cognitive cognitive radio network, where full-duplex SUs encounter both self-interference and co-channel interference. \cite{9127944}  & Use tiny DL-based CSS, which can learn temporal correlation dynamics from the SU's locally sensed information, including the activity pattern of the PUs and self-interference suppression coefficient  \\  \hhline{~--}
 & Resource allocation policy for D2D pairs in a D2D underlay network, aimed at minimizing the mutual interference from neighboring D2D pairs and maximizing the sum data rate of all D2D pairs \cite{kaiUnderlayRA2023}  & The D2D resource allocation problem can be formulated as a multi-agent RL problem, where each D2D pair acts as an RL agent and uses a tiny RL algorithm to devise its resource allocation policy \\
\hline
SEI & SEI resilient to emitter spoofing \cite{9846906}  & RF-based SEI, composed of feature embedding and emitter classification stages, is resistant to emitter spoofing. Here, tiny DL can be utilized to perform the feature embedding  \\
\hline
AGC index &  Make RX resilient to interference signal strength changes during the reception period \cite{joly2024algorithm}  & Utilize tiny DL to derive a range for the AGC index, which would help the RX remain resilient to interference changes during the next reception period \\
\hline
Transmission power management & Transmission power management on BLE beacons to mitigate interference and enhance the navigation system performance \cite{9209253}  & Use tiny RL to continuously tune the transmission power of the beacon, directly on the beacon, by taking into account RSSI of the received signal  \\
\hline
Proximity detection & Classical RSSI-based ranging among mobile devices results in the estimated distance, among devices, with a low accuracy \cite{9371718} & Use tiny DL-cum-RSSI-based ranging, which would increase the accuracy of distance estimate by taking into account the time-domain/frequency-domain/statistical features of RSSI    \\
\hline
\end{tabular}
\end{table*}
}

\subsection{Key Takeaways}

A summary of the challenges and respective research directions in untapped tinyML deployment possibilities is available in Table~\ref{summary_UNtapped_table}.
Key takeaways are as follows.
In near-field MRI-WPT systems, tiny DL may help estimate the system characteristics on the TX side, estimate TX and RX compensation capacitances, and act as the RX load classifier on the TX side, while tiny DRL can enable continuous tuning of the TX side compensation capacitance to ensure constant power at the RX load resistor.
These capabilities can enhance efficiency and autonomy in near-field MRI-WPT systems.
In BA-WSNs and for batteryless devices equipped with WuR, tiny RL can assist resource optimization.
For example, tiny RL can enable the distributive tuning of the reflection coefficient of sensor nodes in BA-WSNs to ameliorate the spectral efficiency at the sink and optimization of the duty cycle of batteryless devices to prolong their sensing lifetime.
In IoT-based dynamic process monitoring systems, tiny RL may enable GoC-based device scheduling, while tiny DL can enable lightweight anomaly detection locally on IoT devices.
These capabilities enable autonomous optimization of data transmission decisions in IoT-based dynamic process monitoring systems.
At the physical layer, tiny DL may help realize efficient and lightweight signal processing capabilities on RXs, such as low-power edge nodes.
Representative signal processing tasks include channel estimation in UVLC, AMC, RF-based SEI, and estimation of the AGC index range of RX.
Executing these signal processing tasks locally may promote faster inference, reduced transmission overhead, and improved RX adaptability.
In cognitive radio networks and indoor navigation systems, tinyML may help optimize resource utilization, such as subcarrier allocation and transmission power tuning, to mitigate interference and localization on mobile RXs.

\begin{figure}[!t]
\centering
\includegraphics[width=\linewidth]{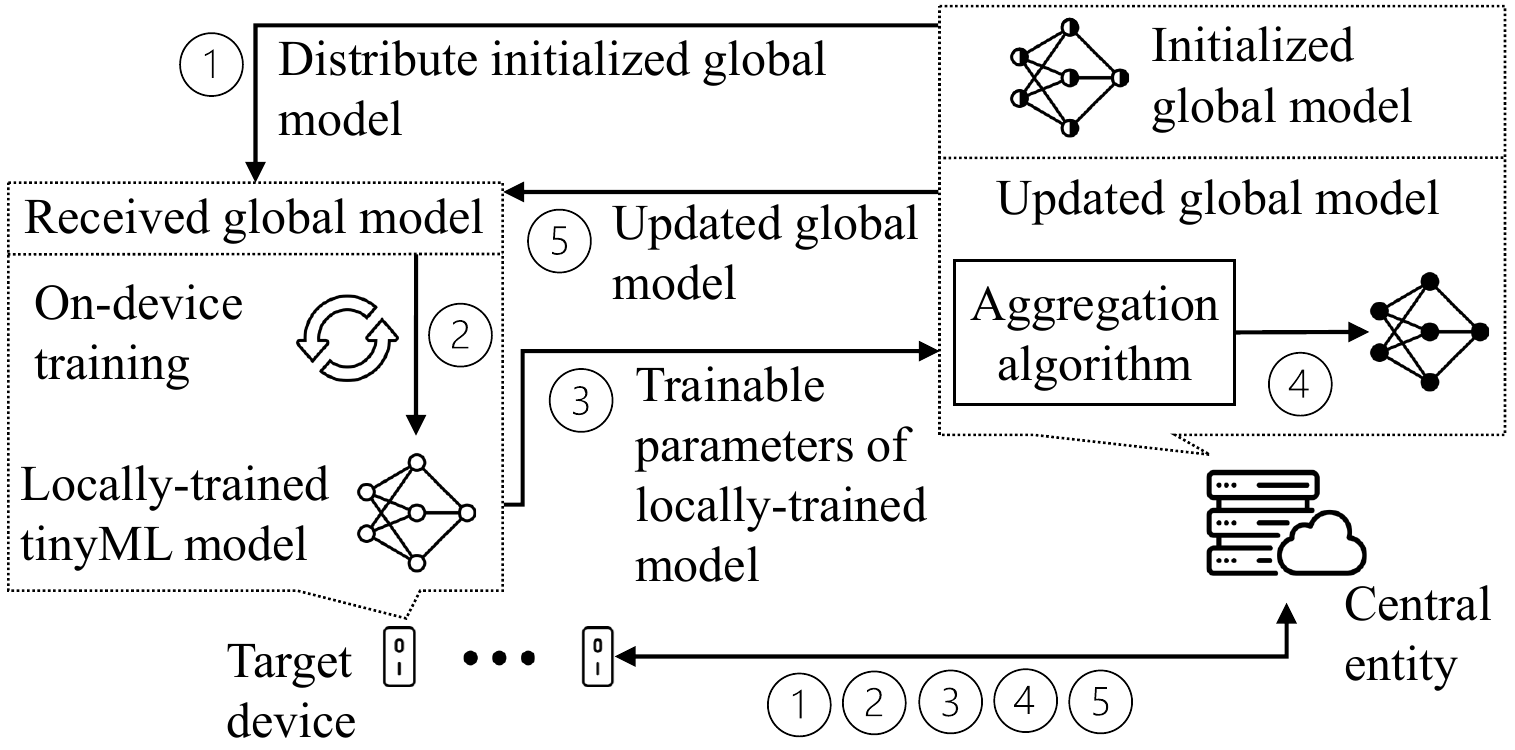}
\caption{Centralized tinyFL-based tinyML model update procedure.
First, the central entity distributes the initialized global tinyML model to the target devices.
Target devices update the received global tinyML model through on-device training, and transmit the trainable parameters of its respective locally-trained tinyML model to the central entity.
After receiving the model parameters sent by target devices, the central entity updates its global tinyML model.
Finally, the updated global tinyML is distributed to the target devices.}
\label{FL_MLtraining_figure}
\end{figure}

\section{Mitigating the Concept Drift}\label{dealConceptDrift}

In mutable-environment systems with devices equipped with an on-device NN-based ML model, the statistical properties of the training data often differ unexpectedly from the data generated during operation.
Such a phenomenon is known as concept drift \cite{10196602}, with different types illustrated in Fig.~\ref{conceptDriftTypesfigure}.
The repercussion of the concept drift is that the patterns learned from the training data are no longer valid for the current data, resulting in poor decisions/predictions from the ML model \cite{8496795}.
Retraining the ML model is an obvious way to grapple with the concept drift \cite{10196602, 10251832}.
Next, we provide approaches to retrain tinyML models available on battery-powered and batteryless devices.

\subsection{Battery-Powered Device Case}\label{FL_batteryPowered_Algos}

Recently, centralized tiny FL (tinyFL), illustrated in Fig.~\ref{FL_MLtraining_figure}, has gained popularity for updating on-device tinyML models.
The centralized tinyFL approach enables tinyML model updates in a collaborative manner \cite{vu2025integration}, reduces the privacy risks as only the ML model's trainable parameters are shared with the central entity \cite{10916688}, and mitigates ML model training issues arising due to the lack of training data on MCU-based devices \cite{fi16110413}.
The corresponding main steps are as follows \cite{vu2025integration}:
\begin{enumerate}
    \item\label{step1_FL} The central entity distributes the initialized global tinyML model to the target devices (MCU-based devices).
    Here the central entity could be an edge/cloud server.
    \item\label{step2_FL} Target devices update the received global tinyML model through on-device training.
    This local training is essential as it enables the on-device tinyML model to learn the device-specific data patterns/insights.
    \item Each target device transmits the trainable parameters of its respective locally-trained tinyML model to the central entity.
    \item After receiving the model parameters sent by target devices, the central entity updates its global tinyML model by minimizing the global loss function.
    This minimization problem is also called the aggregation problem and it takes into account the recently received model parameters of target devices.
    Aggregation algorithms, such as FedAvg \cite{10916688}, TinyReptile \cite{10916688}, proportion weighting strategy \cite{vu2025integration}, etc, can be used to solve the aggregation problem.
    \item\label{step5_FL} Finally, the updated global tinyML is distributed to the target devices.
\end{enumerate}
Step~\ref{step2_FL}-\ref{step5_FL} are repeated until the global loss function is minimized to the desired level.
There is another variant of tinyFL called decentralized tinyFL \cite{fi16110413}, with the downside of adopting computationally expensive paradigms, such as swarm learning \cite{fi16110413}, to exercise collaborative ML model updates.
Due to this, the decentralized tinyFL approach is not favorable for updating tinyML models on MCU-based devices.

As mentioned in Step~\ref{step2_FL} of the centralized tinyFL-based tinyML model update procedure, target devices update the received global tinyML model through on-device training.
However, training an NN-based tinyML model, such as a tiny DL model, with full backpropagation directly on an MCU-based device is challenging, particularly when the device lacks dynamic RAM (DRAM) and has limited memory resources like static RAM (SRAM) and FLASH.
These challenges arise due to the following factors \cite{10284551}:
\begin{itemize}
  \item During training, the calculation of the gradients requires storing intermediate activations.
  This leads to the memory requirements during training being significantly higher than those during inference.
  Given that the available memory on MCUs is often just sufficient for inference, accommodating these additional requirements is particularly challenging.
  \item On-device NN-based ML models are quantized models, while optimizing them is difficult because of mixed-precision tensors.
\end{itemize}
This showcases the need to revamp the backpropagation procedure.
Meanwhile, \cite{10284551} has highlighted the following three methods for revamping backpropagation to allow on-device training.

{
\setlength\arrayrulewidth{1pt}
\begin{table}[!t]
\caption{Accuracy Findings with respect to the Training Phase of the MCUNet-5FPS Model to Different Tensor Precisions \label{gradientScaling_table}}
\centering
\begin{tabular}{@{}p{1.5cm} p{1.4cm} p{1.5cm} l@{}}
\hline
\textbf{Dataset} & \textbf{Fp32} & \textbf{Int8} & \textbf{Int8$+$Gradient Scaling}  \\
\hline
Cars & $56.7\%$ & $31.2\%$ & $55.2\%$  \\ 
Food & $67.1\%$ & $52.5\%$ & $64.4\%$  \\
Flowers & $88.8\%$ & $84.5\%$ & $89.1\%$  \\
CF$10$ & $86\%$ & $75.4\%$ & $86.9\%$  \\
CF$100$ & $63.4\%$ & $54.5\%$ & $64.6\%$  \\
\hline
\end{tabular}
\end{table}
}

\subsubsection{Gradient Scaling}

The quantized NN-based model/graph comprises tensors with mixed precision, including int8, int32, and fp32.
The existence of these multi-precision tensors destabilizes the gradient-update procedure, which makes the optimization/training of the quantized NN-based model challenging.
The gradient-update destabilization can be quantified through the \textit{weight norm-by-gradient norm ratio}, where destabilization is directly proportional to the ratio \cite{10284551}.
Note that quantized weights are approximately the scaled version of the original floating-point weights. As a result, the gradients computed with respect to the quantized weights also become scaled versions of the gradients of the original weights. Because of the scaling factor, the weight norm-by-gradient norm ratio becomes significantly higher for the quantized NN-based model compared to its floating-point counterpart \cite{10284551}. Therefore, to address the gradient-update destabilization problem, we have to explicitly rescale the gradients of the quantized weights. This gradient scaling restores the weight norm-by-gradient norm ratio of the quantized NN-based model to be approximately the same as in its floating-point counterpart. As a result, the gradient scaling resolves the gradient-update destabilization problem and enables the quantized NN-based model's training possible \cite{10284551}.

Now, let us take a look at the accuracy findings, in Table~\ref{gradientScaling_table}, with respect to the training phase of the following three versions of the MCUNet-5FPS model \cite{10284551}: tensors with fp32, tensors with int8, and tensors with int8 plus gradient scaling-based training.
Here, all three model versions perform the classification task on the following datasets: Cars~\cite{10284551}, Food~\cite{10284551}, Flowers~\cite{10284551}, CF$10$~\cite{10284551}, CF$100$~\cite{10284551}.
Table~\ref{gradientScaling_table} shows that the accuracy achieved by the int8 version is significantly less than its fp32 counterpart.
However, the accuracy achieved by the int8-plus-gradient scaling version is comparable to its fp32 counterpart.
This indicates, the gradient scaling also increases the accuracy of a quantized NN-based ML model without any additional memory overhead.

\begin{figure}[!t]
\centering
\includegraphics[width=\linewidth]{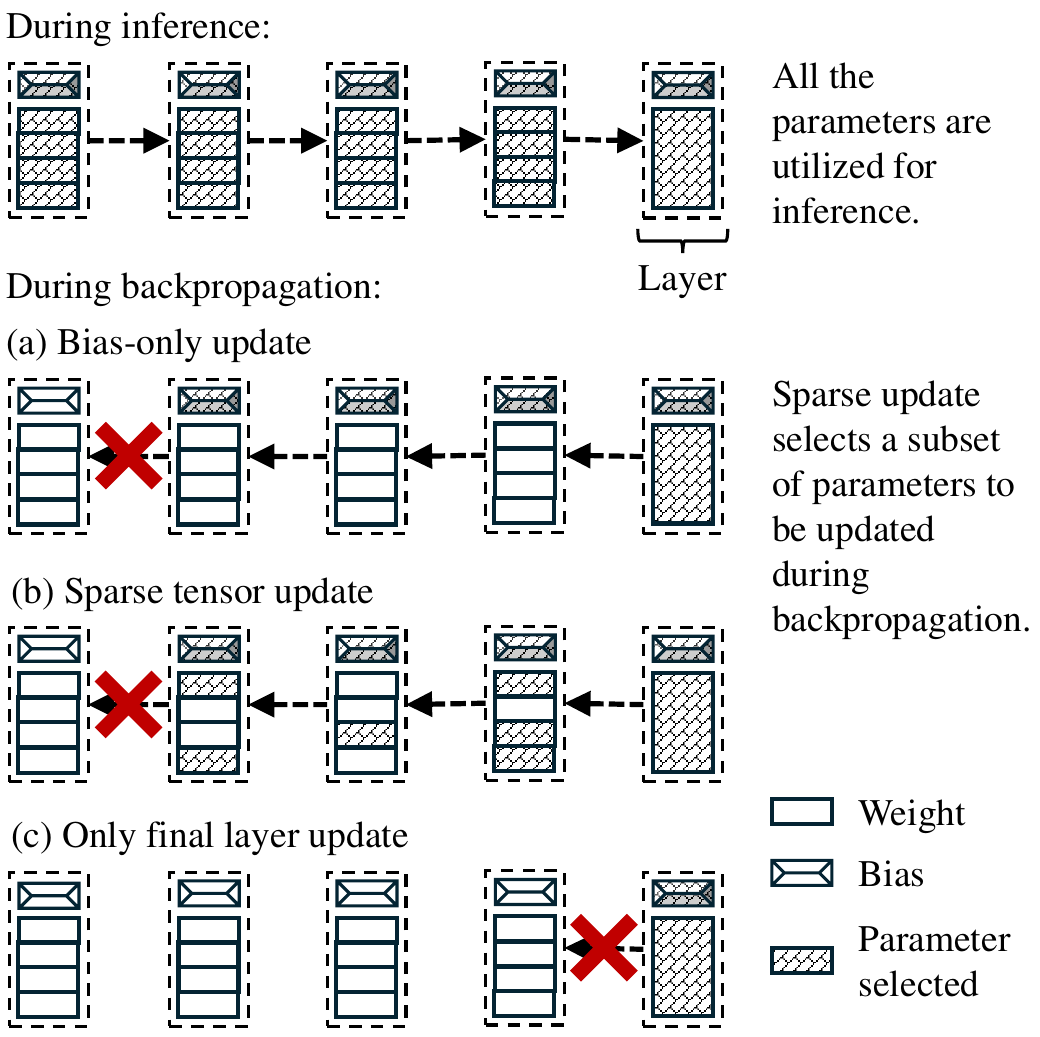}
\caption{Types of sparse update procedures to be used during backpropagation.}
\label{sparseTensorUpdatefigure}
\end{figure}

\subsubsection{Sparse Update}

One effective way to reduce memory footprint during training an NN-based model is to skip updating non-critical parameters of the model by pruning their respective gradients during backpropagation \cite{10284551}.
As illustrated in Fig.~\ref{sparseTensorUpdatefigure}, sparse update during backpropagation consists of:
\begin{itemize}
  \item Sparse tensor update: single out a subset incorporating the information about NN layers whose weights need to be updated.
  Next, for each of the selected layers, single out a subset of weights that need to be updated.
  \item Bias-only update: only the biases from the previously selected NN layers are updated.
  \item Only final layer update: update only the parameters available in the final layer of the NN-based model.
\end{itemize}

Sparse update is carried out through a method called \textit{contribution analysis}.
This method involves solving an optimization problem that seeks to find a subset of NN layers, and their respective weights/biases, to update, such that the model's accuracy is maximized and the model update/training task's memory footprint stays within the available memory limit.
Subsequently, its objective function and constraint take into account a weight's/bias's contribution to the model's accuracy and the model update/training task's memory footprint, respectively \cite{10284551}.
Meanwhile, the objective function quantifies the aforementioned accuracy contributions in the following two ways:
(i) the contribution of performing the bias-only update in the last $k$ NN layers compared to updating only the final layer parameters, and 
(ii) the contribution of performing the sparse tensor update, i.e., update weights, in one more NN layer compared to performing the bias-only update in that respective layer.

Next, let us take a look at the findings, in Table~\ref{sparseUpdate_table}, with respect to the training phase of the MobileNetV2-w0.35 model \cite{10284551}.
Among the four model update methods available in Table~\ref{sparseUpdate_table}, the highest accuracy is achieved in the case of the sparse update method.
Moreover, the \textit{extra memory} utilized in the case of the sparse update method is $\sim3.6\times$ smaller than the \textit{update last $k$ layers} method.
This indicates that the sparse update method increases the model's accuracy as well as reduces the memory footprint of the model updating/training task.

{
\setlength\arrayrulewidth{1pt}
\begin{table}[!t]
\caption{Accuracy and Extra Memory Footprint Findings with respect to the Training Phase of the MobileNetV2-w0.35 Model \label{sparseUpdate_table}}
\centering
\begin{tabular}{@{}l l l l@{}}
\hline
\textbf{Model Update Method} & \textbf{Accuracy}$^{\dagger}$ & \textbf{Extra Memory}$^{\ddagger}$ & \textbf{$k$} \\
\hline
Sparse update & $\sim72\%$ & $\sim138$kB & $-$  \\
Update last $k$ layers & $\sim70\%$ & $\sim545$kB & $36$  \\
Update only biases in last $k$ layers & $\sim64\%$ & $\sim134$kB & $36$  \\
Update only last layer & $\sim60\%$ & $-$ & $1$  \\
\hline
\multicolumn{4}{@{}p{\linewidth}@{}}{$^{\dagger}$Accuracy is the average of accuracies obtained with regard to the following datasets \cite{10284551}: Cars, CF10, CF100, CUB, Flowers, Food, Pets, VWW.}\\
\multicolumn{4}{@{}p{\linewidth}@{}}{$^{\ddagger}$\textit{Extra memory} is defined as the difference between the memory utilized in case-A and case-B.
Here, case-A is the method available in the first column of Table~\ref{sparseUpdate_table}, while case-B is the \textit{update only last layer} method.
Moreover, \textit{extra memory} is computed analytically.}
\end{tabular}
\end{table}
}

\begin{figure}[!t]
\centering
\includegraphics[width=\linewidth]{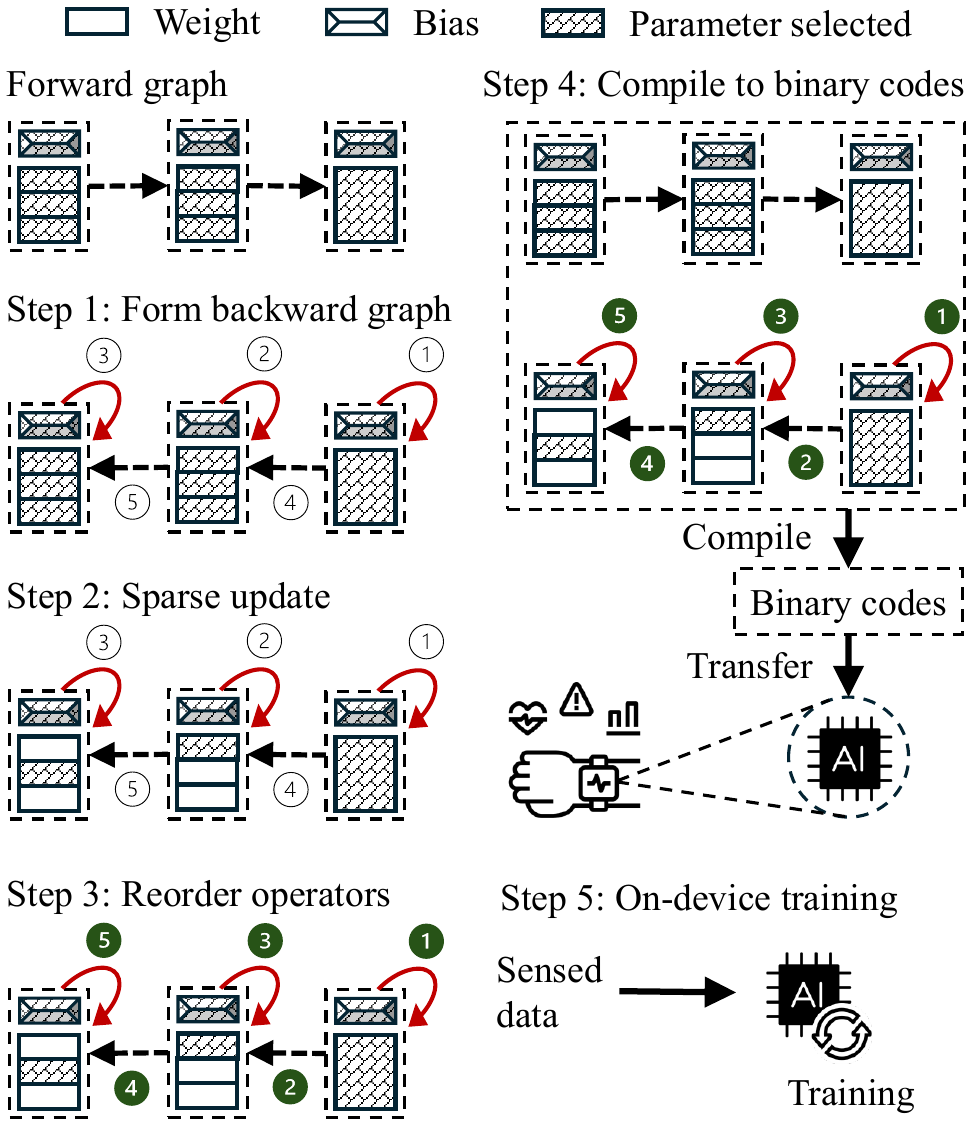}
\caption{Illustration of the working of the tiny training engine.
The red cycles in the figure denote the gradient descent operators in NN layers.
The tiny training engine first generates a static backward graph from the original forward graph.
Next, redundant parameters of the backward graph are pruned using the sparse update.
The tiny training engine then reorders the execution of operators in the pruned backward graph.
Finally, the pruned and reordered backward graph, along with the forward graph, are compiled into binary codes and transferred to the target device.}
\label{TTEfigure}
\end{figure}

\subsubsection{Training Engine}

\textit{Gradient scaling} and \textit{sparse update} alone cannot reduce the memory footprint of the training procedure.
This motivated the authors in \cite{10284551} to design a \textit{tiny training engine}, whose working is illustrated in Fig.~\ref{TTEfigure}.
The tiny training engine first generates a static backward graph from the original forward graph to facilitate on-device training.
Note that the backward graph transfers the differentiation operation, which is typically performed during runtime, to the compilation.
This means that by optimizing the backward graph, we can curb the memory footprint of the model training task.
Thus, in the next step of the tiny training engine, redundant parameters of the backward graph are pruned using the \textit{sparse update}, to reduce the memory footprint of the backward graph.
After pruning the backward graph, the tiny training engine checks the dependence of all tensors, such as weights, gradients, and activations, in the backward graph.
This dependence check allows the tiny training engine to reorder the execution of operators present in the backward graph.
Such reordering facilitates operator fusion, further reducing the backward graph's memory footprint.
Finally, the pruned and reordered backward graph, along with the forward graph, are compiled into binary codes implementable on the target device.

Next, let us take a look at the findings, in Table~\ref{TTE_table}, with respect to the training phase of the MobileNetV2-w0.35 model \cite{10284551}.
Table~\ref{TTE_table} shows that both the training latency and peak memory are the least in the case of the tiny training engine.
Moreover, relative to the full backpropagation, the training latency and peak memory have been reduced by $\sim23\times$ and $\sim21\times$, respectively, in the case of the tiny training engine.

\subsection{Batteryless Device Case}

The centralized tinyFL-based tinyML model update procedure, from Section~\ref{FL_batteryPowered_Algos}, in combination with the \textit{over-the-air aggregation} \cite{10000620} can be used in the case of batteryless devices.
The over-the-air aggregation utilizes the waveform superposition property of multiple-access channels to enable aggregation of the trainable parameters during transmission \cite{10000620, 9791337}.
This is accomplished with the following steps \cite{10000620, bagci2025update}:
(i) the central entity schedules a subset of target devices (batteryless devices) based on their computing capability, available energy, and quality of the channel between the device and the central entity;
(ii) the target devices in the aforementioned subset simultaneously transmit the trainable parameters of their respective locally-trained tinyML model to the central entity; and
(iii) the central entity utilizes multiple receive antennas along with combining techniques to alleviate channel fading and recover the noisy aggregated signal.

{
\setlength\arrayrulewidth{1pt}
\begin{table}[!t]
\caption{Latency and Peak Memory Footprint Findings with respect to the Training Phase of the MobileNetV2-w0.35 Model \label{TTE_table}}
\centering
\begin{tabular}{l l l l@{}}
\hline
\textbf{Method} & \textbf{Latency Per} & \textbf{Measured Peak} & \textbf{Accuracy}$^{\dagger}$  \\
& \textbf{Image} & \textbf{Memory} &  \\
\hline
Tiny training engine & $373$ms & $141$kB & $75.1\%$  \\
Sparse update$^{\circledast}$ & $3448$ms & $335$kB & $75.1\%$  \\
Full backpropagation$^{\circledast}$ & $8501$ms & $2939$kB & $75.1\%$  \\
\hline
\multicolumn{4}{@{}p{\linewidth}@{}}{$^{\circledast}$Sparse update/full backpropagation + Tensorflow Lite micro kernels.}\\
\multicolumn{4}{@{}p{\linewidth}@{}}{$^{\dagger}$Accuracy is the average of accuracies obtained with regard to the following datasets \cite{10284551}: Cars, CF10, CF100, CUB, Flowers, Food, Pets, VWW.}
\end{tabular}
\end{table}
}

In the tinyFL-based procedure from Section~\ref{FL_batteryPowered_Algos}, tinyML model updates occur on an MCU-based device either through on-device training or with respect to the global model.
However, frequent power interruptions/failures are faced by batteryless devices.
Due to this, these model update tasks might have to face issues such as code fragmentation and code corruption, potentially resulting in a non-functional device.
Meanwhile, the \textit{intermittent-aware modify operation} from \cite{10569023} resolves the aforementioned issues related to the intermittent model update.
The intermittent-aware modify operation works as follows.\newline
Consider two code blocks.
One contains the original instructions of a model, while another contains the reformed instructions that will replace a portion of the original ones.
The former is the original code block, while the latter is the reformed code block.
Meanwhile, the original code block's aforementioned portion targeted for replacement is called the replacement block.
Based on the size of the reformed and replacement block, there are three possible cases:
\begin{enumerate}
    \item If the reformed code block is larger than the replacement block, the replacement is done in stages. First, instructions in the replacement block are replaced by as many reformed instructions as the replacement block can accommodate. Additional memory locations are then allocated to store the remaining reformed instructions. Next, the original code block's instruction flow is modified so that execution jumps from the original code block to the newly allocated memory. After the reformed instructions run, execution jumps back to the instruction that follows the replacement block.
    \item\label{step2_modifyOperation} If the reformed code block is of the same size as the replacement block. Then, instructions in the replacement block are replaced by the reformed instructions at the same memory locations.
    \item If the reformed code block is smaller than the replacement block, the replacement is done in steps. First, instructions in the replacement block are replaced by as many reformed instructions as the reformed code block contains. Next, the original code block's instruction flow is modified so that after the reformed instructions run, execution jumps to the instruction that follows the replacement block.
\end{enumerate}
Note that when the tinyML model update occurs through on-device training, only case~\ref{step2_modifyOperation} is possible.
The reason is that only weights/biases get updated, not operations incorporating these weights/biases, in the case of on-device training.

Checkpointing is frequently applied in intermittent computing platforms/devices to periodically save the ongoing task's state such that the task can be resumed from the saved state after an unexpected interruption.
Traditional checkpointing techniques are only suitable for routine tasks like inference, while the \textit{update-aware checkpointing} technique from \cite{10569023} facilitates checkpointing for both the routine and model update tasks.
For the latter, the device's SRAM is divided into two regions: \textit{stack}, used during routine tasks, and \textit{update buffer}, used to hold {the aforementioned reformed code block.
Upon the occurrence of a low-power interrupt, the \textit{stack} and all CPU registers are backed up to create the routine task checkpoint, while the \textit{update buffer} and CPU registers, specific to the model update task, are backed up to create the update task checkpoint.

Traditionally, bootloaders responsible for managing the tinyML model update task do not support checkpointing \cite{10569023}.
Meanwhile, the \textit{fault-tolerant bootloader} from \cite{10569023} supports checkpointing for both the routine and model update tasks, and comprises a fault recovery mechanism to resolve the code fragmentation issue.
The fault-tolerant bootloader performs the model update task without a system reboot, which is particularly useful for the case: 
when the size of the aforementioned reformed code block is significantly larger than the one supported by the batteryless device's supercapacitor.
Then, the reformed code block has to be split into several small code blocks.
In this case, multiple small model updates, one for each small code block, are performed to complete the model update task.
That means a significant amount of device energy will be consumed if a system reboot is performed after every small model update.
Meanwhile, the fault-tolerant bootloader's no-reboot operating method saves device energy by preventing the reboot-based energy consumption.

Next, let us compare two model update methods: FLoRa \cite{10_11453583120_3586963} (baseline) and the framework from \cite{10569023}.
FLoRa supports the aforementioned reformed code block's split into several small code blocks.
However, unlike the framework from \cite{10569023}, FLoRa does not perform small model updates on the aforementioned small code blocks.
Next, FLoRa demands the whole reformed code block must be stored in the device's SRAM to advance towards the model update task.
Consequently, the model update task would fail if the size of the reformed code block is larger than the one supported by the supercapacitor.
Moreover, FLoRa has no mechanism to backup the reformed code block, which is already stored in the device's SRAM, upon the occurrence of a low-power interrupt.
Consequently, the whole reformed code block must be recollected in the device's SRAM to restart the model update task following a low-power interrupt.
Thus, relative to the framework from \cite{10569023}, the energy consumed to restart the model update task after a low-power interrupt should be higher in FLoRa.
Findings from \cite{10569023}, with respect to the LeNet-5 Model, prove this deduction as FLoRa requires $\sim67.5$mJ of energy to restart the model update task, while the framework from \cite{10569023} requires $\sim2.5$mJ.
Additionally, the framework from \cite{10569023} achieves a finite total model update time of $\sim1350$ms in the case of a $10$mF capacitor, while FLoRa does not.
This indicates that FLoRa fails to complete the model update task.

\subsection{Key Takeaways}

Key takeaways are as follows.
The centralized tinyFL-based tinyML model update procedure and backpropagation revamping methods, named gradient scaling, sparse update, and training engine, can be used for both device cases.
The centralized tinyFL approach enables tinyML model updates in a collaborative manner, reduces privacy risks, and mitigates ML model training issues arising due to the lack of training data on devices.
The gradient scaling method resolves the gradient-update destabilization problem and enables the quantized NN-based model's training possible.
The sparse update method seeks to find a subset of NN layers, and their respective weights/biases, to update, such that the model's accuracy is maximized and the model update/training task's memory footprint stays within the available memory limit.
The training engine follows a series of steps, illustrated in Fig.~\ref{TTEfigure}, to reduce the memory footprint and latency of the model update/training task.
In the batteryless device case, the centralized tinyFL-based tinyML model update procedure should be used in combination with over-the-air aggregation, intermittent-aware modify operation, update-aware checkpointing, and the fault-tolerant bootloader.
The over-the-air aggregation utilizes the waveform superposition property of multiple-access channels to enable aggregation of the trainable parameters during transmission.
The intermittent-aware modify operation resolves issues, such as code fragmentation and code corruption, related to the intermittent model update.
The update-aware checkpointing technique facilitates checkpointing for both the routine and model update tasks.
Lastly, the fault-tolerant bootloader supports checkpointing for both the routine and model update tasks, and comprises a fault recovery mechanism to resolve the code fragmentation issue.

\section{Conclusion}\label{conclUSion}

This article provided insights into the utilization of tinyML-based end-device/edge inference within wireless networks.
Specifically, we discussed the existing frameworks accustomed to design tinyML algorithms, the tapped/untapped deployment possibilities of tinyML in the wireless networks, and the tinyML model update procedure for battery-powered/batteryless end-devices.
Our key takeaways are:
\begin{itemize}
  \item Existing literature on wireless networks has primarily leveraged tinyML, specifically tiny DL, for feature extraction-based tasks, as summarized in Table~\ref{summary_EXISTING_table}.
  In contrast, tiny RL, suited for decision-making tasks in wireless networks, remains unexplored.
  \item TinyML has vast deployment possibilities within the realms available in Table~\ref{summary_UNtapped_table}.
  \item The centralized FL-based tinyML model update approach increases privacy during model update and mitigates model training issues arising due to the lack of training data on devices.
  \item Backpropagation revamping methods, namely gradient scaling, sparse update, and tiny training engine, for on-device training increases the accuracy of a quantized tinyML model as well as reduces the memory footprint and latency of the model updating/training task.
  \item Methods, namely update-aware checkpointing, fault-tolerant bootloader, and intermittent-aware modify operation, not just resolve issues faced by the model update task on batteryless end-devices, but also reduce the energy consumed to restart the model update task.
\end{itemize}

To conclude, Table~\ref{PAIT6G_KRD_table} provides key challenges and research directions with respect to the protocol aspects for integration of tinyML to 6G wireless networks. Specifically, Table~\ref{PAIT6G_KRD_table} discusses the tinyML-largeML integration, datasets for performance evaluation of tinyML-based systems, energy lifecycle profiling of tinyML-based systems, trustworthy and interpretable decision-making, and sociotechnical challenges of embedded AI.

{
\setlength\arrayrulewidth{1pt}
\begin{table*}[!t]
\caption{Protocol Aspects for Integration of TinyML to 6G Wireless Networks: Key Challenges and Research Directions\label{PAIT6G_KRD_table}}
\centering
\begin{tabular}{@{}p{1.45cm} p{8cm} p{8cm}@{}}
\hline
\textbf{Realm} & \textbf{Challenges} & \textbf{Research Direction} \\ 
\hline
TinyML-largeML integration & The integration of tinyML and largeML in 6G presents several challenges, including synchronization of tinyML's and largeML's inference, interoperability issues, etc \cite{vu2025integration}  & Devise standardized frameworks that define task-specific roles and interoperability requirements for workflow management, data coordination, and runtime compatibility among tinyML and largeML models  \\
\hline
Datasets & Current datasets primarily capture static or low-mobility scenarios, limiting the performance evaluation of adaptive tinyML approaches, such as online learning, incremental model updates, and context transfer mechanisms, essential for seamless mobility \cite{11370394}  & Devise datasets for high-speed mobility scenarios, such as vehicular ($>120$ km/h) and aerial ($>200$ km/h). Note that these datasets must capture mobility traces, handover events, and channel coherence time variations to enable the performance evaluation of adaptive tinyML approaches   \\  \hhline{~--}
 & Datasets needed for performance evaluation of tinyML models against 6G targets, such as peak data rates, connection density, and energy efficiency, are not available \cite{11370394}  & Devise datasets, along with respective specifications to be captured in them, to evaluate the performance of tinyML models against 6G targets and accelerate the integration of tinyML in 6G edge environments  \\
\hline
Energy lifecycle profiling & Energy lifecycle pipeline of tinyML-based systems composed of several stages, including device wake-up schedule, data sensing, data preprocessing,$^{\S}$ inference, communication, and battery self-discharge. Omitting any of the pipeline stage(s) leads to unrealistic assumptions about battery life and sustainability achieved by tinyML-based systems \cite{11266886}  & Devise an end-to-end energy lifecycle modeling framework that comprehensively captures all the pipeline stages  \\
\hline
Ethical integration & Sociotechnical challenges, including consent acquisition in passive monitoring systems, compliance with data minimization regulations, algorithmic bias, and ambiguous liability in the event of system failures, significantly influence the adoption or rejection of tinyML solutions in 6G \cite{11266886}  & Incorporate privacy preserving mechanisms, including on-device anonymization, zero-retention buffers, and differential privacy guards \cite{11266886}, at the firmware level and devise standardized guidelines for embedded AI in 6G  \\
\hline
Trustworthy and interpretable decision-making & Safety-critical fields, including healthcare and autonomous transportation, demand trustworthy and interpretable decision-making from tinyML. However, interpretability methods, such as the Shapley additive explanation \cite{minh2022explainable} and local interpretable model-agnostic explanation \cite{minh2022explainable}, are computationally expensive and require post-hoc processing \cite{11266886}  & Integrate proxy methods, such as rule-extraction techniques \cite{Li2025AsurveyNT}, for interpretability into the tinyML inference pipeline. Next, utilize lightweight validation method from \cite{aldughaim2020incremental} to introduce trustworthiness in real-time decision-making \\
\hline
\multicolumn{3}{@{}p{17cm}@{}}{$^{\S}$ Depending on data and application, preprocessing involves operations such as fast Fourier transform, Mel-spectrogram generation \cite{11434719}, discrete wavelet transform filtering \cite{10029140}, data labeling \cite{10029140}, high-/low-pass spectral filtering \cite{10113053}, etc.}\\
\end{tabular}
\end{table*}
}

\bibliographystyle{IEEEtran}
\bibliography{IEEEabrv,references}

% \newpage

 

% \end{IEEEbiography}



% \vfill

\end{document}